\pdfoutput=1
\documentclass[12pt]{article}
\usepackage{graphicx}
\usepackage{subcaption}
\usepackage{amssymb}
\usepackage{amsmath}
\usepackage{amsfonts}
\usepackage{multirow}
\usepackage{xspace}

\usepackage{xcolor}
\usepackage{url}
\usepackage{cancel}
\usepackage{ulem}
\usepackage{float}
\usepackage{float}

\usepackage{pifont}% http://ctan.org/pkg/pifont

\usepackage{soul}
\usepackage[english]{babel}
\usepackage[T1]{fontenc}
\usepackage{lmodern}
\usepackage[utf8]{inputenc}  
\usepackage{bbm}

\newcommand{\bwp}{wino/bino{\scriptsize (+)}\xspace}
\newcommand{\bwm}{wino/bino{\scriptsize ($-$)}\xspace}
\newcommand{\him}{higgsino{\scriptsize ($-$)}\xspace}

\newcommand{\sihi}{singlino/higgsino\xspace}

\usepackage{hyperref}
\hypersetup{
  colorlinks   = true, %Colours links instead of ugly boxes
  urlcolor     = blue, %Colour for external hyperlinks
  linkcolor    = black, %Colour of internal links
  citecolor   = black %Colour of citations
}

\usepackage[numbers,sort&compress]{natbib}
\graphicspath{{plots/}}

\newcommand{\lsim}
{\;\raisebox{-.3em}{$\stackrel{\displaystyle <}{\sim}$}\;}
\newcommand{\gsim}
{\;\raisebox{-.3em}{$\stackrel{\displaystyle >}{\sim}$}\;}

\newcommand\tb{\tan\beta}
\newcommand{\ctb}{\cot\beta}

\newcommand\SB{s_\beta}
\newcommand\CB{c_\beta}

\newcommand\ReDiag{\mathop{%
  \raise .5pt\hbox{[}%
  \widetilde{\mathrm{Re}}%
  \raise .5pt\hbox{]}}}
\newcommand\ReOffDiag{\mathop{%
  \raise .5pt\hbox{$\llbracket$}%
  \widetilde{\mathrm{Re}}%
  \raise .5pt\hbox{$\rrbracket$}}}

\newcommand\SW{s_\mathrm{w}}
\newcommand\CW{c_\mathrm{w}}
\newcommand\MW{M_W}
\newcommand\MZ{M_Z}

\newcommand\Sl{\tilde l}

\newcommand\msl[1]{m_{\Sl_{#1}}}

\newcommand\gl{{\tilde g}}
\newcommand\mgl{m_\gl}

\newcommand\ino[1]{\tilde\chi_{#1}}

\newcommand\chapm[1]{\ino{#1}^\pm}

\newcommand\cha{\chapm}
\newcommand\mcha[1]{m_{\chapm{#1}}}

\newcommand\neu[1]{\ino{#1}^0}
\newcommand\mneu[1]{m_{\neu{#1}}}

\newcommand\refeq[1]{Eq.~(\ref{#1})}

\newcommand\refse[1]{Sect.~\ref{#1}}

\newcommand\citere[1]{Ref.~\cite{#1}}
\newcommand\citeres[1]{Refs.~\cite{#1}}

\newcommand{\CP}{{\cal CP}}
\newcommand{\cp}{{\CP}}

\newcommand{\tev}{\,\, \mathrm{TeV}}
\newcommand{\gev}{\,\, \mathrm{GeV}}

\newcommand{\He}{h_1}
\newcommand{\Hz}{h_2}
\newcommand{\Hd}{h_3}
\newcommand{\Ae}{A_1}
\newcommand{\Az}{A_2}
\newcommand{\mHe}{m_{\He}}
\newcommand{\mHz}{m_{\Hz}}
\newcommand{\mAz}{m_{\Az}}
\newcommand{\mHp}{m_{H^\pm}}

\newcommand{\ETmiss}{\ensuremath{E_T\hspace{-4.5mm}\slash}\hspace{2.5mm}}

\newcommand\MO{\texttt{MicrOMEGAs}}

\newcommand\HB{\texttt{HiggsBounds}}
\newcommand\HS{\texttt{HiggsSignals}}
\newcommand\HT{\texttt{HiggsTools}}

\newcommand\pb{\ensuremath{\,\mbox{pb}}}

\newcommand{\sig}{\sigma}

\def\order#1{\ensuremath{{\cal O}(#1)}}
\def\reffi#1{\mbox{Fig.~\ref{#1}}}

\def\ga{\gamma}
\def\De{\Delta}

\def\la{\lambda}
\def\ka{\kappa}
\newcommand{\mueff}{\mu_{\mathrm{eff}}}

\def\gmin2{\ensuremath{(g-2)_\mu}}

\newcommand{\ssi}{\ensuremath{\sig_{\rm SI}^p}}
\newcommand{\ssdp}{\ensuremath{\sig_{\rm SD}^p}}
\newcommand{\ssdn}{\ensuremath{\sig_{\rm SD}^n}}
\newcommand{\Och}{\Omega_\chi h^2}

\definecolor{Orange}{named}{orange}
\definecolor{Purple}{named}{purple}
\definecolor{Lightblue}{cmyk}{0.9,0.1,0.1,0.3}
\definecolor{dgelborange}{cmyk}{0.,0.3,0.5, 0.}
\definecolor{Lila}{rgb}{0.5,0.,1}
\definecolor{Darkgreen}{rgb}{0.,.7,0.2}

\evensidemargin \oddsidemargin
\allowdisplaybreaks
\begin{document}
\thispagestyle{empty}

\def\thefootnote{\fnsymbol{footnote}}

\begin{flushright}
\mbox{}
IFT--UAM/CSIC--25-157 %, 
%arXiv:2512.nnnnn [hep-ph]
\end{flushright}

\begin{center}

{\large\sc 
{\bf Consistent Excesses in the LHC Electroweak SUSY Searches:\\[.5em]
GUT-based Singlino/Higgsino Interpretation in the NMSSM}}

\vspace{0.3cm}

{\sc
Emanuele Bagnaschi$^{1}$%
\footnote{email: emanuele.angelo.bagnaschi@lnf.infn.it}
, Manimala Chakraborti$^{2}$%
\footnote{email: M.Chakraborti@soton.ac.uk}% 
, Sven Heinemeyer$^{3}$%
\footnote{email: Sven.Heinemeyer@cern.ch}%
\mbox{}\\[.1cm] and Ipsita Saha$^{4}$%
\footnote{email: ipsita@iitm.ac.in}
}

\vspace*{.5cm}

{\sl
$^1$INFN, Laboratori Nazionali di Frascati, Via E.\ Fermi 40,
00044 Frascati (RM), Italy

\vspace{0.1cm}

$^2$School of Physics and Astronomy, University of Southampton,
Southampton, SO17 1BJ,\\
United Kingdom (former address)

\vspace*{0.1cm}

$^3$Instituto de F\'isica Te\'orica (UAM/CSIC), 
Universidad Aut\'onoma de Madrid, \\ 
Cantoblanco, 28049, Madrid, Spain

\vspace*{0.1cm}

$^4$Department of Physics, Indian Institute of Technology Madras, Chennai 600036, India
}

\end{center}

\vspace*{0.1cm}

\begin{abstract}
\noindent
The search for signatures of supersymmetric models remains one of the
main item on the BSM search program at the LHC, with EW SUSY partners still
allowed with masses as low as a few hundred GeV.
Over the last years, searches for the ``golden channel'',
$pp \to \neu2 \cha1 \to \neu1 Z^{(*)} \, \neu1 W^{\pm (*)}$ show consistent
excesses between ATLAS and CMS in the 2~soft-lepton and 3~soft-lepton 
plus \ETmiss\ searches, assuming $\mneu2 \approx \mcha1 \gsim 200 \gev$ and
$\De m_{21} := \mneu2 - \mneu1 \approx 20 \gev$.
We interpret these excesses in the framework of the Next-to-Minimal
Supersymmetric Standard Model (NMSSM). We assume a singlino dominated
lightest neutralino, $\neu1$,
as a Dark Matter (DM) candidate. The second and third lightest
neutralinos, $\neu{2,3}$ are higgsino like, with the higgsino mixing
parameter $\mu$ being smaller than the soft SUSY-breaking bino and
wino masses, $M_1$ and $M_2$. We furthermore assume the approximate
GUT relations $M_1 \sim M_2/2 \sim M_3/6$, with the implication for our scenario
of a gluino mass $\mgl \sim M_3 \gsim 3 \tev$.
Scalar masses are assumed to heavy and do not play a role in our analysis.
We find that this scenario is in agreement with all relevant
experimental constraints, comprising the LHC searches for SUSY
particles and additional Higgs bosons, the LHC Higgs-boson rate
measurements, the DM direct detection limits and the upper limit on the 
DM relic density.
We demonstrate that this scenario gives an excellent description of the
observed excesses in the search for 2~and 3~soft-leptons plus \ETmiss,
with $\mneu2 \sim \mneu3 \sim \mcha1$ and $\De m_{21} \sim 20 \gev$.
This constitutes the first explanation of the soft-lepton excesses in
a model with GUT relations among the soft SUSY-breaking 
parameters. 

\end{abstract}

%\pacs{}

\def\thefootnote{\arabic{footnote}}
\setcounter{page}{0}
\setcounter{footnote}{0}

\newpage

%%%%%%%%%%%%%%%%%%%%%%%%%%%%%%%%%%%%%%%%%%%%%%%%%%%%%%%%%%%%%%%%%%%%%%%%%%%%%%%
%%%%%%%%%%%%%%%%%%%%%%%%%%%%%%%%%%%%%%%%%%%%%%%%%%%%%%%%%%%%%%%%%%%%%%%%%%%%%%%

\section{Introduction}
\label{sec:intro}
The Standard Model (SM) of particle physics remains one of our most successful theories ever,
being able to describe with high precision (most of) 
the wealth of currently available experimental data, covering a large span in energy scales.
Yet we know from cosmological and astrophysical observations that physics beyond the SM
is required to explain various phenomena, such as the ones interpreted today in terms of Dark Matter (DM).
One of the most promising models, among the plethora of theories Beyond the SM (BSM)
containing a viable DM candidate, is the Minimal Supersymmetric (SUSY) Standard Model (MSSM)~\cite{Ni1984,Ba1988,HaK85,GuH86} (for a recent review, see \citere{Heinemeyer:2022anz}).
The MSSM features, for each SM field, a superpartner that differs by one-half unit spin.
Moreover, supersymmetry requires extending the Higgs sector from one to two Higgs doublets.
As a consequence, the physical spectrum of the theory contains five Higgs bosons
instead of a single one, as instead happens in the SM.
The Higgs sector therefore consists of the light and heavy $\cp$-even Higgs bosons
($h$ and $H$), the $\cp$-odd Higgs boson ($A$), as well as a pair of
charged Higgs bosons ($H^\pm$).
Concerning the electroweak (EW) sector of the MSSM, it contains the
SUSY partners of the SM leptons, the scalar leptons (sleptons), as
well as the charged (neutral) SUSY partners of the charged (neutral)
EW gauge bosons (gauginos) and Higgses (higgsinos). After electroweak symmetry breaking,
gauginos and higgsinos mix to form two charginos ($\cha{1,2}$) and four neutralinos ($\neu{1,2,3,4}$),
the so-called ``EWinos''. The searches for the production and decays of the EWinos
and sleptons in various final states are one of the primary targets in
BSM searches at the LHC.

In the MSSM with conserved R-parity (which we assume throughout the
paper), the lightest neutralino, $\neu1$, assumed to be the lightest
SUSY particle (LSP), can play the role of a weakly interacting massive particle
(WIMP) DM candidate~\cite{Go1983,ElHaNaOlSr1984}.
The masses and mixings of the neutralinos and charginos are governed
by two soft SUSY-breaking parameters, the $U(1)$ parameter $M_1$,
the $SU(2)$ parameter $M_2$, as well as the SUSY-conserving Higgs mixing parameter~$\mu$.
The latter parameter poses a theoretical problem. It is a SUSY-conserving
parameter, but its value is not protected by any symmetry. However,
experimentally it must be at the scale of the other soft SUSY-breaking
parameters. This was dubbed the~``$\mu$-problem'' of the
MSSM~\cite{Kim:1983dt}. 

In order to solve the ``$\mu$-problem'', the extension of the~MSSM by a gauge-singlet superfield -- 
the Next-to MSSM (NMSSM)~\cite{Maniatis:2009re,Ellwanger:2009dp} -- was introduced~\cite{Kim:1983dt}.
The addition of the singlet superfield, containing a singlet scalar and a singlet fermion (the singlino)
enlarges the spectrum of the theory.
The neutralino sector of the NMSSM, with the addition of the singlino,
contains five neutralinos, $\neu{1,2,3,4,5}$.
The Higgs-boson spectrum is extended by one $\cp$-even and one $\cp$-odd
Higgs-boson, resulting in three $\cp$-even states ($\He$, $\Hz$, $\Hd$),
two $\cp$-odd states ($\Ae$, $\Az$) and a pair of charged Higgs-bosons
($H^\pm$). Here we identify the $\He$ with the Higgs boson discovered
at the LHC with $\mHe \sim 125 \gev$. 

In addition to the MSSM parameters, the NMSSM has additional independent parameters
related to the singlet superfield.
Following the common nomenclature, we choose: $\la$, $\ka$, $A_\ka$, $A_\la$ 
and the effective parameter $\mu_{\mathrm{eff}}$ as independent inputs.
In the NMSSM, the $\mu$~parameter of the MSSM is generated dynamically
as $\mueff := \la\,v_s$ (where $v_s$ denotes the singlet vev), solving the above mentioned $\mu$~problem.
From this relation one can determine the value of $v_s$ as a function of $\lambda$ and $\mu_{\mathrm{eff}}$.

One of the most effective approaches to probe the EW (N)MSSM parameter space,
is to search for the lighter EWinos, $\neu{2,3}$ and $\cha1$, 
in final states containing two or three (possibly soft) leptons accompanied by
substantial missing transverse energy ($\ETmiss$)~\cite{ATLAS-SUSY,CMS-SUSY,Dicus:1983cb,Chamseddine:1983eg,Baer:1985at,Baer:1986vf,Baer:1986dv}. 
These searches turn out to be particularly challenging in the region of
parameter space where the mass difference 
between the initially produced and the final state EWinos are
small~\cite{Buanes:2022wgm,Carpenter:2023agq}, making 
the visible decay products, i.e.\ the leptons, to be rather
soft~\cite{Baer:1994nr,Baer:2011ec,Abdughani:2019wss,Bagnaschi:2017tru,GAMBIT:2018gjo}.
The mass configuration is commonly referred to as ``compressed
spectra''. The CMS and ATLAS collaborations are actively searching for
the EWinos in this compressed spectra region.
Interestingly, over the last years searches for the ``golden
channel'' (focusing for simplicity to the MSSM particle content), 
$pp \to \neu2 \cha1 \to \neu1 Z^{*} \, \neu1 W^{\pm *}$ show consistent
excesses between ATLAS and CMS in the 2~soft-lepton and 3~soft-lepton
plus missing energy~\cite{ATLAS:2019lng,CMS:2021edw}. This also holds for
the combined 2/3~soft-leptons plus \ETmiss~\cite{ATLAS:2021moa,CMS:2024gyw,CMS:2025ttk}
searches assuming $\mneu2 \approx \mcha1$ for a mass difference of 
$\De m_{21} := \mneu2 - \mneu1 \approx 20 \gev$.%
\footnote{In the following we denote as $\De m_{ij} := m_i - m_j$, where $i,j = 1,2,3$ stands for the three lightest neutralinos, 
and $i,j = \pm$ for the lightest chargino.}
Therefore, it seems naturally interesting to identify
the underlying parameter configuration of the EW (N)MSSM and the associated
properties of the EWinos that can give rise to such excesses at the LHC.

In \citere{CHS6} within the MSSM three different scenarios were analyzed,
classified by the hierarchy and signs of the EWino mass parametes,
$M_1$, $M_2$ and $\mu$, while assuming heavy sleptons.
These scenarios are named after their DM candidate, i.e.\ the largest
component of the $\neu1$. The mass configuration favored by the
experimental excesses in the compressed 
spectra searches for EWkinos is naturally found in two of these scenarios:
\begin{itemize}
\item[(i)]
\bwp\ DM with $\cha1$-coannihilation ($M_1 \lsim M_2 < \mu$),
with $M_1 \times \mu > 0$, 
\item[(ii)]
\bwm\ DM with $\cha1$-coannihilation ($|M_1| \lsim M_2 < \mu$), 
with $M_1 \times \mu < 0$, 
\item[(iii)]
\him\ DM ($|\mu| < |M_1|, M_2 $), 
with $M_1 \times \mu < 0$\,.
\end{itemize}
In \citere{CHS6} all relevant experimental constraints were taken into
account.  The experimental results comprised  of 
the DM relic abundance~\cite{Planck},
the DM direct detection (DD) experiments~\cite{LZ-new} and
the direct searches at the LHC~\cite{ATLAS-SUSY,CMS-SUSY}.%
\footnote{
~A wider range of this type of models taking into account these
constraints was analyzed previously in
\citeres{CHS1,CHS2,CHS3,CHS4,gmin2-mw,CHS5}.
}%
~In \citere{CHS6} it was found that the \bwm\ scenario gives the best
description of the excesses, followed by \bwp, which tended to
slightly lower $\De m_{21}$. On the other hand, it was concluded that
the \him\ scenario cannot yield a sufficiently large $\De m_{21}$. This was
particularly due to the bounds from DD experiments, excluding the
range $\De m_{21} \gsim 10 \gev$, which would, however, be required to
describe the soft-lepton excesses. 

Other analyses of the soft-lepton excesses interpreted within the (N)MSSM framework can be found in \citeres{Ellwanger:2024vvs,Agin:2024yfs,Martin:2024pxx,Agin:2025vgn,Hammad:2025wst,Araz:2025bww,Constantin:2025bqp}.
More in detail, in \citeres{Martin:2024pxx,Agin:2024yfs,Agin:2025vgn,Araz:2025bww}, various MSSM scenarios have been explored in the 
context of the soft-lepton excesses. While \citere{Martin:2024pxx} investigated the higgsino-like LSP scenario within the MSSM, 
\citeres{Agin:2024yfs,Agin:2025vgn,Araz:2025bww} examined both the mono-jet and soft-lepton excesses simultaneously for a  simplified higgsino-like scenario. Another combination of various experimental searches has recently been published  in \citere{Constantin:2025bqp}.
Within the NMSSM framework, \citere{Ellwanger:2024vvs} considered a light higgsino-like triplet corresponding to the lightest chargino and 
next-to-lightest neutralinos, while assuming a singlino-like LSP that is nearly mass-degenerate with the triplet states. This study also 
imposed an additional constraint requiring a 95~GeV singlet-like Higgs state.
In contrast, \citere{Agin:2024yfs} assumed a higgsino–singlino 
mass splitting of 5–20~GeV, where the decay products from the higgsino-to-singlino transition contribute to
the observed signal. The aim of 
that analysis was to identify the best-fit parameter region capable of accommodating both the soft-lepton and mono-jet excesses. This 
study has also incorporated a recasting of the relevant analysis using the package \texttt{HACKANALYSIS~2}~\cite{Goodsell:2024aig}. More recently, 
\citere{Hammad:2025wst} performed a comprehensive parameter scan of the NMSSM parameter space, where the soft-lepton excess along with 
a 95~GeV and a 650~GeV scalar resonance was also discussed. 

As noted in ~\citere{CHS6}, the \bwm\ and \bwp\ scenarios, while giving a good
description of the soft-lepton excesses and complying with the DM
experimental constraints, require $|M_1| \sim |M_2|$. Here it is
important to note that
one of the most compelling arguments in favor of SUSY is the
unification of forces at the Grand Unified Theory (GUT)
scale~\cite{GUT-unification}. In simple GUT realizations, such as the
CMSSM~\cite{cmssm}, this yield the following mass pattern at the EW
scale: $M_1 \sim M_2/2 \sim M_3/6$, where $M_3$ is the $SU(3)$ soft
SUSY-breaking parameter, which coincides with the gluino mass,
$M_3 \sim \mgl$. Searches for colored particles at the LHC have set a
limit of $\mgl \gsim 2 \tev$~\cite{CMS:2019ybf,ATLAS:2024lda}, for the relatively light $\neu1$ we are interested in,  leading to $|M_1| \gsim 350 \gev$, which is too large to yield $\mneu1 \lsim 200 \gev$.
Consequently, the MSSM scenarios that can describe the soft-lepton excesses do not follow the GUT-based mass pattern. On the other hand, a GUT-based mass pattern seems to yield too high neutralino and chargino masses, well above $300 \gev$.

In this paper we propose a scenario that can describe well the
soft-lepton excesses, follows the GUT-based mass patterns and is in
agreement with the searches for gluinos: the NMSSM with a singlino
dominated LSP, with $200 \gev \approx |\mu| < M_1 \sim M_2/2 \sim M_3/6$, 
and $\mgl \gsim 2 \tev$. This parameter space corresponds to
higgsino-like $\neu{2,3}$ and $\cha1$. The LHC searches are
interpreted as searches for
$pp \to \neu{2,3} \cha1 \to \neu1 Z^{*} \, \neu1 W^{\pm *}$ in the
NMSSM, leading to the desired 2~or 3~soft-lepton plus \ETmiss\ final
states. We show that this \mbox{``\sihi''} scenario can give rise to the `correct'
abundance of DM relic density, while agreeing with the latest DD
bounds. 
%We also find that future DD experiments will not be able to conclusively test this scenario due to part of the preferred parameter space been below the neutrino floor.
%
Concerning the DD limits, the exchange of all $\cp$-even Higgs bosons
can contribute in a significant way, see \citere{Arganda:2025fhx} for a recent
analysis. Consequently, we leave the 
Higgs-boson sector parameters as free parameters, leading to a variation
of the new Higgs-boson mass scales and the properties of the $\He$.
We demonstrate that besides the above listed characteristics of the
chargino/neutralino sector also the Higgs-boson sector properties are
in agreement with the LHC searches for BSM Higgs bosons, as well as
with the current LHC Higgs-boson rate measurements.
However, the possibility is left open to probe deviations with respect to the
SM in the properties of the lightest $\cp$-even Higgs at $\sim 125 \gev$,
or the existence of new Higgs bosons, at future runs of the (HL-)LHC.

This paper is organized as follows. In \refse{sec:model} we briefly review
the parameters of the chargino/neutralino and Higgs sectors of NMSSM, fix our
notations and define the scenario under investigation.
The relevant constraints for this analysis, in particular the excesses
in the searches for $\neu{2,3} \cha1$ at the LHC, are briefly summarized
in \refse{sec:constraints}. The scenario we are analyzing, together
with our analysis flow is detailed in \refse{sec:paraana}. 
In \refse{sec:results} we present the details of our results as well as
the prospects for future DD experiments.
The conclusions can be found in \refse{sec:conclusion}.

%%%%%%%%%%%%%%%%%%%%%%%%%%%%%%%%%%%%%%%%%%%%%%%%%%%%%%%%%%%%%%%%%%%%
%%%%%%%%%%%%%%%%%%%%%%%%%%%%%%%%%%%%%%%%%%%%%%%%%%%%%%%%%%%%%%%%%%%%

\section {The chargino/neutralino and Higgs sectors of the NMSSM}
\label{sec:model}

In this section,  we review the chargino/neutralino and the Higgs sectors of the NMSSM, in order
to highlight the most important properties of the model, and fix our notation.
The scalar quark and lepton sectors are assumed to be heavy 
and not to play a relevant role in our analysis. Throughout this
paper we also assume that all parameters are real, i.e.\ the absence of
$\CP$-violation.

We consider the $Z_3$ invariant NMSSM with the superpotential
\begin{align}
\label{Vsup}
W_{\mathrm{NMSSM}} &= 
  \la\,\hat S \, {\hat H}_2 \cdot {\hat H}_1
  \,+\, \frac{\ka}{3} \, {\hat S}^3\, + \ldots\,.
\end{align}
Here the ellipses denote the Yukawa couplings of the Higgs
superfields, which are the same as in the MSSM. Denoting the vev of the scalar
component of the singlet superfield as $\langle S \rangle = v_s$, see \refeq{Hdecomp} below, 
the first term in the superpotential gives rise to the effective
$\mu$~term,
\begin{align}
\mueff := \la\,v_s\,.
\label{eq:mueff}
\end{align}
In the remainder of the paper we will omit the subscript $_{\mathrm{eff}}$. 
We furthermore denote with $\tb$ the ratio of the two
vevs of the two Higgs doublets of MSSM, $\tb = v_2/v_1$.

%%%%%%%%%%%%%%%%%%%%%%%%%%%%%%%%%%%%%%%%%%%%%%%%%%%%%%%%%%%%%%%%%%%%

\subsection{The chargino/neutralino sector}

The NMSSM neutralinos are the linear superpositions of the
neutral $SU(2)_L$ and $U(1)_Y$ gauginos, neutral higgsinos and the singlino
$\tilde B, \tilde {W^3}$, $\tilde H_2^{0}$, $\tilde H_1^{0}$ and
$\tilde S$, respectively.
Their masses and mixings are determined by $U(1)_Y$ and $SU(2)_L$
gaugino masses $M_1$ and $M_2$, the Higgs mixing parameter $\mu$
and the singlino mass term. 
The neutralino mass matrix in the basis
$(-i \tilde B, -i \tilde W^3, \tilde H_2^{0}, \tilde H_1^{0}, \tilde S )$
is given by
\begin{equation}
M_{N}=\left(
\begin{array}{ccccc}
  {M_1} & 0 & -\MZ \CB \SW & \MZ \SB \SW & 0 \\
  0 &   {M_2} & \MZ \CB \CW   & -\MZ \SB \CW  & 0 \\
  -\MZ \CB \SW & \MZ \CB \CW  & 0 &   {-\mu} & -\la v_2 \\
  \MZ \SB \SW & -\MZ \SB \CW  &   {-\mu} & 0 & -\la v_1 \\
  0 & 0 & -\la v_2 & -\la v_1 & 2 \ka v_s
\end{array} \right)\,,
\end{equation}
where $\CB~(\SB)$ denotes $\cos\beta~(\sin\beta)$
and $\CW = \sqrt{1 - \SW^2} = \MW/\MZ$ denotes effective weak
leptonic mixing angle, and $\MW, \MZ$ are the mass of the
$W$~and $Z$~boson, respectively. After diagonalization, the five eigenvalues
of the matrix give the five neutralino masses
$\mneu1 < \mneu2 < \mneu3 < \mneu4 < \mneu5$. The diagonalization
matrix is denoted as $N$, such that 
\begin{equation}
N \, M_N \, N^T = \mathrm{diag}(\mneu1, \mneu2, \mneu3, \mneu4, \mneu5)\,.
\end{equation}
As discussed above, the lightest neutralino, $\neu1$ is the LSP and
is assumed to give rise to part or the full DM relic density,
see \refse{sec:constraints} below.

The chargino mass eigenstates, as in the MSSM, are given by the mixing
between the charged winos and higgsinos $(\tilde {W^\pm}, \tilde H_{2/1}^{\pm})$
respectively with their mass matrix given by,
\begin{equation}
M_{C}=\left(
\begin{array}{cc}
 {M_2} & \sqrt 2 \MW\CB \\\sqrt 2 \MW\SB &  {\mu }\,.
\end{array} \right)
\end{equation}
Diagonalizing $M_C$ with a bi-unitary transformation, yields the two
chargino-mass eigenvalues $\mcha1 < \mcha2$.

Throughout our analysis we neglect $\cp$-violation and
assume $\mu, M_1, M_2$ to be real.
From general considerations, even sticking to real parameters, one
could choose some (or all) of the mass parameters negative. However,
here it is important to note that the results for physical observable
are affected 
only by certain combinations of signs (or, in general, phases).
Without loss of generality, it is possible to rotate the phase of
$M_2$ away, i.e.\ choose ($M_2$ real and) $M_2 > 0$.
This leaves in principle the signs of $M_1$ and $\mu$ free.
We choose $M_1$ to be positive, to allow for the GUT condition
$M_1 \sim M_2/2 \sim M_3/6$, leaving the sign of $\mu$ free. 
Consequently, we will explore both combinations,
$(M_1 \times \mu) > 0$ and $(M_1 \times \mu) < 0$.

%%%%%%%%%%%%%%%%%%%%%%%%%%%%%%%%%%%%%%%%%%%%%%%%%%%%%%%%%%%%%%%%%%%%%%%%%%

\subsection{The Higgs-boson sector}

Following \refeq{Vsup} the Higgs potential contains additional terms w.r.t.\ the MSSM case, 
\begin{align}
V_{\mathrm{NMSSM}} \supset | \la (H_1^+ H_2^- \,-\, H_1^0 H_2^0) + \ka S^2 |^2
  \,+\, (\la A_\la (H_1^+ H_2^- \,-\, H_1^0 H_2^0)S \,+\, \ka A_\ka S^3 \,+\, \mathrm{h.c.})\,,
\end{align}
where $H_1$, $H_2$ and $S$ denote the scalar parts of their respective superfields.
Assuming the following decomposition for the two neutral components of the doublets and the one singlet Higgs fields, 
\begin{align}
\label{Hdecomp}
H^0_{i = 1,2} = v_i + \frac{1}{\sqrt{2}} \left( S_i + i P_i \right), \qquad \qquad S = v_s + \frac{1}{\sqrt{2}} \left(S_3 + i P_3 \right)\,,
\end{align}
the mass matrix for the $\cp$-even sector is given by,\\[2em]

\begin{align}
&\mathcal{M}_{\cp-\mathrm{even}} = \nonumber \\ &= \begin{pmatrix} 
\bar{g}^2 v_1^2 - \mu\tb A_{\Sigma} & (2 \lambda^2 - \bar{g}^2) v_1 v_2 + \mu A_{\Sigma} & 2 \lambda\mu v_1 + \lambda v_2 (A_{\Sigma} + \kappa v_s )\\
(2 \lambda^2 - \bar{g}^2) v_1 v_2 + \mu A_{\Sigma} & \bar{g}^2 v_2^2 - \mu \ctb A_{\Sigma} & 2 \lambda\mu v_2 + \lambda v_1 (A_{\Sigma} + \kappa v_s) \\
2 \lambda\mu v_1 + \lambda v_2 \left(A_{\Sigma} + \kappa v_s \right) & 2 \lambda\mu v_2 + \lambda v_1 \left(A_{\Sigma} + \kappa v_s \right) & -\lambda A_{\lambda } \frac{v_1 v_2}{v_s} + \kappa v_s \left(A_{\kappa} + 4 \kappa v_s \right)
\end{pmatrix}
\end{align}
where we have followed the notation of \citere{Slavich:2013bqa}, and defined $A_\Sigma = A_\lambda + \kappa v_s$  and $\bar{g}^2 = (g^2+g'^2)/2$. Upon diagonalization by an orthogonal matrix $R^s$, we have three $\cp$-even mass-ordered eigenstates $h_{1,2,3}$. Note that the tree-level upper bound on the lightest $\cp$-even mass is higher than in the MSSM and given by:
\begin{align}
m^2_{h_1} < m_Z^2 \cos^2 \left(2\beta\right) + \lambda^2 (v_1^2+v_2^2) \sin^2 \left( 2 \beta \right).
\end{align}
The additional term is relevant for values of $\tb \lsim 10$. 

The mass matrix for the $\cp$-odd sector is given by: 
\begin{align}
&\mathcal{M}_{\cp-\mathrm{odd}} = \nonumber \\ &= \begin{pmatrix} 
- \mu\tb A_{\Sigma} & - \mu A_{\Sigma} & -\lambda v_2 \left(A_{\Sigma} - 3 \kappa v_s \right) \\
-\mu A_{\Sigma} & - \mu\ctb A_{\Sigma} & -\lambda v_1 \left(A_{\Sigma} - 3 \kappa v_s \right) \\
- \lambda v_2 \left( A_{\Sigma} - 3 \kappa v_s \right) & -\lambda v_1 \left( A_{\Sigma} - 3 \kappa v_s \right) & -4 \lambda \kappa v_1 v_2 - \lambda A_{\lambda} \frac{v_1 v_2}{v_s} - 3 \kappa A_{\kappa} v_s
\end{pmatrix}.
\end{align}
After diagonalization via an orthogonal matrix, we obtain three eigenstates, $G^0$, $A_1$ and $A_2$, where the first one is the
would-be Goldstone boson that gives mass to the $Z$, and $A_1$ and $A_2$ are two $\cp$-odd Higgses, again with mass-ordered labeling.

On the top of the neutral Higgses, we have a charged Higgs boson sector with the same structure as in the MSSM.
Thus, the physical spectrum of the Higgs sector of the NMSSM contains 
three $\cp$-even states ($\He$, $\Hz$, $\Hd$, ordered in mass),
two $\cp$-odd states ($\Ae$, $\Az$, ordered in mass) and a pair of
charged Higgs-bosons ($H^\pm$).

We use the code {\tt NMSSMTools}~\cite{Ellwanger:2004xm,Ellwanger:2005dv,Domingo:2015qaa,NMSSMTOOLS-www} as our spectrum generator,  which we instruct to compute the Higgs sector masses using the full 1-loop and the 2-loop corrections computed in \citere{Degrassi:2009yq}, and implemented in the code.
The following set of parameters is used as input:
$\lambda$, $\kappa$, $A_{\lambda}$, $A_{\kappa}$, $\mu_{\mathrm{eff}}$, with $v_s$ obtained from \refeq{eq:mueff}, 
to define the NMSSM Higgs sector, on top what is already required in the MSSM.

We identify the lightest $\cp$-even Higgs boson, $\He$, with the Higgs
boson discovered at the LHC with $\mHe \sim 125 \gev$.
Since the loop corrections involving the top/stop sector are of particular importance for the NMSSM Higgs-boson sector
(see \citere{Slavich:2020zjv} for a review), for the diagonal soft SUSY-breaking parameters in the stop/sbottom sector  
we have chosen fixed values that are sufficiently large 
to make it easy to achieve $\mHe \sim 125 \gev$~\cite{Slavich:2020zjv}, i.e.~$M_{Q_3} = M_{U_3} = M_{D_3} = 3 \tev$.
Moreover, with this choice the masses of the scalar tops/bottoms are safely above the LHC limits, and we are not 
required to implement stop and sbottom searches in our framework.

While in general it is possible to identify also $\Hz$ with the state at $\sim 125 \gev$ and have $\He$ as a so far undetected Higgs-boson (see, however,
\citeres{Domingo:2018uim,Choi:2019jts,Biekotter:2021qbc,Ellwanger:2024vvs,Choi:2019yrv,Cao:2019ofo,Li:2022etb,Ellwanger:2023zjc,Cao:2024axg,Ellwanger:2024txc,Lian:2024smg} as well as 
\citeres{Biekotter:2017xmf,Biekotter:2019gtq,Hollik:2018yek} for closely related SUSY models),  
we leave this possibility for future work.

Following the analysis in \citere{CHS6}, but contrary to our previous
analyses~\cite{CHS1,CHS2,CHS3,CHS4,gmin2-mw,CHS5}, the heavy $\cp$-even Higgs bosons can play a relevant role in the
(cancellation of the) contributions to the DD cross sections, where all $\He$, $\Hz$ and~ $\Hd$~exchanges can contribute.

The fact that we treat the Higgs-boson sector parameters as free will
require us to check the Higgs sector against experimental constraints from BSM
Higgs-boson searches, as well as LHC Higgs-boson rate measurements,
as will be discussed below.

%%%%%%%%%%%%%%%%%%%%%%%%%%%%%%%%%%%%%%%%%%%%%%%%%%%%%%%%%%%%%%%%%%%%%%%%%%
%%%%%%%%%%%%%%%%%%%%%%%%%%%%%%%%%%%%%%%%%%%%%%%%%%%%%%%%%%%%%%%%%%%%%%%%%%

\section {Experimental constraints}
\label{sec:constraints}

Here we briefly list the experimental constraints that we apply to our
data sets. These are a combinations of constraints included in {\tt NMSSMTools}, with others that are handled in our code (i.e.~DM constraints, properties of $H_{125}$ and
the constraints on the extended scalar sector). As we will describe more in detail in \refse{sec:paraana}, we use {\tt NMSSMTools} as our spectrum generator, but it also includes 
the implementation of some of the constraints that we consider in our analysis.

\begin{itemize}

\item Vacuum stability constraints:\\
In the scenario we are interested in studying (see Sec.~\ref{sec:paraana}) all the sleptons are relatively heavy, and all the corresponding trilinear couplings set to zero. Thus the scalar potential is not influenced in any relevant way by the slepton sector.
In the scalar quark sector, all the soft SUSY-breaking masses are heavy as well.
For stop/sbottom sector, we set $A_b = 0$ while we scan over $A_t$. However,
the combination of heavy mass terms with the Higgs mass constraint makes vacuum stability a non-issue, as it was 
explicitly verified with
the constraint on the stability of the effective potential computed by {\tt NMSSMTools}.

\item Landau poles:\\
{\tt NMSSMTools} checks for the presence of Landau poles in the running couplings of the theory up to $M_{\mathrm{GUT}}$.

\item
DM relic density constraints:\\
The latest results from the Planck experiment~\cite{Planck} provide the
experimental data. The relic density is given by, 
\begin{align}
\Omega_{\rm CDM} h^2 \; = \; 0.120\,  \pm 0.001 \, , 
\label{OmegaCDM}
\end{align}
which we use as an upper bound (calculated using the central value
with 2$\sigma$ upper limit), 
\begin{align}
\Omega_{\rm CDM} h^2 \; \le \; 0.122 \, . 
\label{OmegaCDMlim}
\end{align}
The calculation of the relic density in the NMSSM is performed with
the code  \MO{\tt-6.0}~\cite{Belanger:2001fz,Belanger:2006is,Belanger:2007zz,Belanger:2013oya,Belanger:2014vza,Barducci:2016pcb,Belanger:2018ccd,Belanger:2020gnr,Alguero:2023zol}  as included in {\tt NMSSMTools}.

\item
DM direct detection (DD) constraints:\\
We use the latest spin-independent (SI)
DM scattering cross-section ($\ssi$) limits from the PandaX-4T~\cite{PandaX:2024qfu} experiment that we implement ourselves. 
These results were the strongest published results at the moment of our data evaluation. Meanwhile, also new LZ results~\cite{LZ:2024zvo}
were published, yielding even stronger limits. We will comment on their possible effects (and include them) in future evaluations. 
Also XENONnT published new results~\cite{XENON:2025vwd}, which, however, are weaker than the ones from PandaX-4T.
Limits on spin-dependent (SD) scattering cross sections are again implemented using the PandaX-4T results~\cite{PandaX:2024qfu},
since they were the strongest available, for both SD scattering on protons (SDp) and on neutrons (SDn). 
As will be shown below, only the limits from SDn will play a role, and only a minor one in comparison with the SI limits.
The theoretical predictions are calculated with the code \MO, again as included in {\tt NMSSMTools}. 

For ``underabundant'' parameter points (i.e.\ with $\Omega_{\tilde \chi} h^2 \; \le \; 0.118$), the DM scattering cross-section is rescaled by ($\Omega_{\tilde \chi} h^2$/0.118).  In this way it is taken into account that the $\neu1$ provides only a
fraction of the total DM relic density of the universe.

\item Indirect DM detection:\\
Another set of potentially relevant constraints is given by the indirect
detection (ID) of DM. We do not, however, impose these constraints
on our parameter space because of the well-known large uncertainties
associated with astrophysical factors (e.g.\ the galactic DM density profile, 
theoretical corrections etc., see~\citeres{Slatyer:2017sev,Hryczuk:2019nql,Rinchiuso:2020skh,Co:2021ion}). 
The currently most precise indirect detection limits arise from DM-rich
dwarf spheroidal galaxies, where the uncertainties on the cross section
limits are found in the range of $\sim 2-3$~\cite{McDaniel:2023bju,Fermi-LAT:2015att}. 
The most severe constraint from the latest analysis
in \citere{McDaniel:2023bju}, 
assuming only one single dominant DM annihilation mode, sets a limit
of $\mneu1 \gsim 100 \gev$, for a generic thermal relic
saturating \refeq{OmegaCDM}. However, this limit is much weaker than all
other experimental constraints (in particular the DD limits
described above and the LHC search limits described below)
considered in this study.

\item Constraints from LHC Higgs-boson rate measurements:\\
Any viable BSM model has to accommodate a Higgs boson 
with mass and signal strengths as they were measured at the LHC.
For each point in our parameter space we test (see the next section)
the compatibility of the lightest $\cp$-even scalar in the NMSSM,
$h_1$, with a mass of $\sim 125 \gev$, with the measurements of signal
strengths at the LHC is tested with the code
\HS\,\texttt{v.3}~\cite{Bechtle:2013xfa,Bechtle:2014ewa,Bechtle:2020uwn,Bahl:2022igd}, 
which is included in the code \HT~\cite{Bahl:2022igd,HTnew}.
The code provides a statistical $\chi^2$ for the $h_1$ (or any Higgs
boson assumed to correspond to the state discovered at the LHC)
predictions in comparison to the measurements of the
Higgs-boson signal rates and masses from the LHC.
We consider acceptable parameter points that do not deviate more than
$\De\chi^2 := \chi^2_{h_1} - \chi^2_{\mathrm{min}} < 6.18$,
where the value of $\chi^2_{\mathrm{min}} \simeq \chi^2_{\mathrm{SM}} \approx 151$ is determined from our sample and found to be compatible with a SM-like Higgs.

One possible source for large effects beyond the SM are
additional contributions in the loop-induced process
$h \to \ga\ga$. Here, in particular, a light chargino can yield large
corrections. However, this is automatically tested and taken into
account via the above detailed approach. Conversely, deviations in 
$h \to \ga \ga$ may be interesting targets to cross-check the NMSSM explanation
of the observed excesses. We will comment more on this in Sec.~\ref{sec:results}.
 
\item Constraints from direct Higgs-boson searches at the LHC:\\
The exclusion limits at the $95\%$ C.L.\
of all relevant BSM Higgs boson searches (including Run~2 data from the
LHC  are taken into account with the help of the public code
\HB\,\texttt{v.6}~\cite{Bechtle:2008jh,Bechtle:2011sb,Bechtle:2013wla,Bechtle:2015pma,Bechtle:2020pkv,Bahl:2022igd}, which is included in the public code \HT~\cite{Bahl:2022igd,HTnew}.
~For a parameter point in a particular model (the NMSSM in our case),
\HB\ determines on the basis of expected exclusion limits which is the most
sensitive channel to test each BSM Higgs boson.
In the next step, based on this most sensitive channel, \HB\
determines whether the point is allowed or not at the $95\%$~CL. 

\item Constraints from flavor physics:\\
Flavor constraints can be particularly sensitive to
contributions involving charged Higgs bosons~\cite{Enomoto:2015wbn,Arbey:2017gmh}. 
In particular, the decays $B \to X_s \ga$ and $B_s \to \mu^+ \mu^-$
are most relevant. For the scenarios that we are considering, we do not expect large contributions  to these observables due to the charged Higgs mass being larger than 1 TeV.
We also include other flavor constraints, such as
$\Delta M_d$, $\Delta M_s$, $B^+\to \tau \nu_\tau$, $\Upsilon(1S) \to H/A \gamma$, $m_{\eta_b(1S)}$, $\mathrm{BR}(B \to X_s \mu^+ \mu^-)$, $b \to d \gamma$, $B_d \to \mu^+ \mu^-$, $b \to s \nu \bar{\nu}$ and $K \to \pi \nu \bar{\nu}$, as implemented in {\tt NMSSMTools}. However, they do not have any relevant impact on our analysis.
Moreover, we choose not to consider $(g-2)_{\mu}$ as anomalous. Given our choice of keeping the slepton sector heavy, SUSY contributions to this observables will be very small and therefore automatically compatible with a SM-like measurement.

\item Searches for EWinos at LEP:\\
The constraints on $\mcha1$ and $\mneu1$  (from the measurement of the invisible $Z$-boson decay width) coming from LEP analyses are implemented via {\tt NMSSMTools}. The constraint on the lightest chargino mass, $\mcha1 \gtrsim 100$ GeV, lead to a lower limit of $|\mu| \gsim 100 \gev$, for our scenario. 

\item Searches for EWinos at the LHC:\\
In this analysis we are interested in the $\neu{2,3}$-$\cha1$ pair production
searches with decays via $Z^{(*)}$ and $W^{(*)}$ into final
states involving $2$-$3$ soft leptons and large $\ETmiss$.
Here it should be kept in mind that the original analyses by ATLAS and
CMS were performed in the context of the MSSM, i.e.\ targeting the
process $pp \to \neu2 \cha1 \to \neu1 Z^*\, \neu1 W^*$.
The relevant ATLAS analyses are given
in \citeres{ATLAS:2019lng,ATLAS:2021moa}, whereas the corresponding
CMS analyses are given in \citeres{CMS:2024gyw,CMS:2025ttk}.
However, as discussed in the introduction, in \citere{Agin:2024yfs} a
reinterpretation of the original ATLAS 
data in the context of
a \sihi NMSSM scenario has been performed. We use their
exclusion lines as given in Fig.~8 of \citere{Agin:2024yfs} to exclude
low mass points from our data sample. The relevant characteristics of
our data sample and theirs, particularly the singlino content of the
LSP of 90\% or larger, see the next section, are identical. 
Here it should be stressed
again that the analysis presented in \citere{Agin:2024yfs} did not
include GUT-based values for the fermionic soft SUSY-breaking
parameters, and hence our choices in this respect differ from
theirs. However, this has no impact on the applicability of their
limits on our data set, as $M_1 \sim M_2/2$ are much larger than $\mu$
and play only a minor role in the composition of the $\neu{1,2,3}$ and
$\cha1$. 

As a further check, we passed the points that fulfill all constraints,
including the above described soft-lepton excesses through the code
\texttt{SModelS-v3.1.0}~\cite{Altakach:2024jwk}. We found that
none of these points is excluded by 
\texttt{SModelS}~\cite{Kraml:2013mwa,Ambrogi:2018ujg,Alguero:2020grj,Alguero:2021dig,Altakach:2024jwk}, as expected
from the fact that in the investigated parameter region the most relevant searches remain those focused on the 'compressed' EWino spectra, which lead exactly to the soft-lepton excesses. In this compressed mass region, searches for chargino pair production or associated chargino-neutralino production - resulting in two- or three high $p_T$ leptons plus missing energy signals - are not as pertinent. On the other hand, since the slepton masses are fixed at $1 \tev$ in our analysis, see below, they are not constrained by direct slepton pair production, which would typically produce a two-lepton plus $\ETmiss$ signal.

\end{itemize}

%%%%%%%%%%%%%%%%%%%%%%%%%%%%%%%%%%%%%%%%%%%%%%%%%%%%%%%%%%%%%%%%%%%%%%%%%%
%%%%%%%%%%%%%%%%%%%%%%%%%%%%%%%%%%%%%%%%%%%%%%%%%%%%%%%%%%%%%%%%%%%%%%%%%%

\section{Parameter scan and flow of the analysis}
\label{sec:paraana}

We employ the code \texttt{NMSSMTools} ~\cite{Ellwanger:2004xm,Ellwanger:2005dv,Domingo:2015qaa,NMSSMTOOLS-www} as our spectrum generator%
\footnote{\texttt{NMSSMTools} contains code based on {\tt HDECAY}~\cite{Djouadi:1997yw,Djouadi:2018xqq} 
and {\tt SDECAY}~\cite{Muhlleitner:2003vg}.}%
. It  also computes the branching ratios (BRs) of the Higgs bosons, the Higgs effective couplings, and the 
BRs of the superpartners.

We scan the neutralino/chargino NMSSM parameter space, aiming at fully covering
the regions of low mass neutralinos and charginos consistent with our chosen scenario.
We also scan the Higgs and the scalar-top sector parameter space, where the Higgs sector is relevant for the DD constraints, 
and the stop sector higher-order corrections ensure that we find $m_{h_1} \sim 125 \gev$.
The sleptons are assumed to be sufficiently heavy such that they do
not play any relevant role in our analysis. 
We allow for two sign combinations of $(\mu \times M_1)$ and $A_{\lambda}$, while keeping $M_1$ positive to achieve the GUT conditions for $M_{1,2,3}$. In fact, we use $M_2$ as our input parameter for the gaugino sector, and let {\tt NMSSMTools} impose the GUT conditions on the gaugino soft SUSY-breaking masses.
Our \sihi scenario is defined as follows:

\noindent
{\bf \boldmath{\sihi:} singlino DM with higgsino
dominated \boldmath{$\neu{2,3}$, $\cha1$}:}
\begin{align}
  1000 \gev \leq 2 \times M_1 &= M_2 = M_3/3 \leq 2000 \gev \;, \quad
  150 \gev < | \mu | < 300 \gev \;, \notag\\
  \quad 1000 \gev &= \msl{L} = \msl{R} \;, 
  \quad A_e = A_{\mu} = A_{\tau} = 0\;, \notag\\
  5000 \gev &= M_{\tilde{Q}_{1,2}} = M_{\tilde{u}_R, \tilde{d}_R, \tilde{c}_R, \tilde{s}_R} \;, \quad A_{u,d,c,s} = 0\;, \notag \\
  3000 \gev = M_{\tilde{Q}_3} &= M_{\tilde{t}_R, \tilde{b}_R} \;, \quad A_{b} = 0 \;, \quad 5000 \gev < A_t < 8000 \gev\;, \notag \\
  -0.7 &\le \kappa < 0.7 \;, \quad 0 < \lambda < 5 \;, \quad 0.5 \le \tan\beta \le 50 \notag\;,\\
  5 \gev &\le \left| A_{\lambda} \right| \le 5000 \gev \;, \quad -1000 \gev \le A_{\kappa} \le 1000 \gev\,.
  \label{sihip}
\end{align}
Here $\msl{L,R}$ are the diagonal soft SUSY-breaking parameters in the slepton sector, identical for all three generations,
with $A_{e,\mu,\tau}$ the corresponding trilinear couplings. 
$M_{\tilde{Q}_{1,2}}, M_{\tilde{u}_R, \tilde{d}_R, \tilde{c}_R, \tilde{s}_R}$ denote the diagonal soft SUSY-breaking parameters in the 
scalar quark sector of the first two generations, whereas $M_{\tilde{Q}_3}, M_{\tilde{t}_R, \tilde{b}_R}$ are for the third generations. 
The respective trilinear couplings are $A_{u,d,c,s}$ and $A_{t,b}$. All the  $\overline{\mathrm{DR}}$ parameters are defined at 3 TeV.

%%%%%%%%%%%%%%%%%%%%%%%%%%%%%%%%%%%%%%%%%%%%%%%%%%%%%%%%%%%%%%%%%%%%%%%%%%%%%%%

\bigskip

The data sample is generated by an in-house code that interfaces {\tt NMSSMTools} with {\tt MultiNest}~\cite{Feroz:2008xx}, 
and that implements an ad-hoc likelihood, which includes most of the constraints described above, in order to  drive the sampling algorithm
into the phenomenologically acceptable regions. Spectra corresponding to parameter space points that are either compatible with the
constraints, or not far from the allowed region, are saved into SLHA files~\cite{Skands:2003cj,Allanach:2008qq} for further processing.% 
\footnote{We decide to save also points that are not allowed, albeit not all of them because
of the limited resources available to us, in order to better understand how our code was exploring the parameter space.}

Moreover, in order to sample thoroughly and efficiently our scenario, we split the parameter range into hypercubes, as
required by {\tt MultiNest}, for each one of which we launch a separate instance of our sampling code. We also sampled multiple times each hypercube in order to increase the coverage of our sample.

As a second step, the SLHA files are further processed by running each point through {\tt HiggsTools}, in order to check whether it satisfies or not the constraints coming from the characterization of $h_1$ (with $m_{h_1} \sim 125 \gev$), 
and the ones coming from searches for the Higgses of the extended scalar sector. The "allowed flag" from {\tt HiggsBounds} and the $\chi^2$ from {\tt HiggsSignals} are saved into a custom {\tt BLOCK} inside the SLHA file.

The processed SLHA files, which now also include the information on the compatibility with the
Higgs sector constraints, are saved into {\tt HDF5} archives for efficient storage and analysis.

As a last step, our analysis code processes these {\tt HDF5} files, imposing the definitive set of constraints in a step-wise fashion to create the various point populations that we use to study the phenomenology of the scenario, and that are described in details in the next section.
This applies in particular to the searches for the compressed spectra EWino searches, taken over from \citere{Agin:2024yfs}.

%%%%%%%%%%%%%%%%%%%%%%%%%%%%%%%%%%%%%%%%%%%%%%%%%%%%%%%%%%%%%%%%%%%%%%%%%%

\section{Results}
\label{sec:results}

We follow the analysis flow as described above
and indicate the points surviving certain constraints
with different colors:
\begin{itemize}
\item grey: all scan points that were saved at the sampling level; note that this includes
points that are excluded by current constraints, although not all of them; however, it allows us to identify more clearly the region that was sampled by our code;
\item blue: points that comply with the relic density constraints;
\item cyan: points that additionally pass the DD constraints,
see \refse{sec:constraints};
\item green: points that additionally satisfy the constraints implemented in {\tt NMSSMTools};
\item purple: points that additionally pass \HB;
\item pale orange: points that additionally pass \HS;
\item bright orange: points for which additionally $10 \gev < m_{\tilde{\chi}^0_2} - m_{\tilde{\chi}^0_1} < 35 \gev $;
\item red: points that additionally pass the LHC constraints from searches for $pp \to \neu{2,3} \cha1$ with small $\De m_{21}$, while at the same time being candidate for the excesses that appear in the same searches.
\end{itemize}

We perform our analysis by projecting our multidimensional sample onto two-dimensional planes.
Since the point density has no physical meaning by itself, 
but rather reflects the sampling dynamics, we bin the two-dimensional plane and color each cell according to the 
scheme above if the there is at least one point contained in it. 

%%%%%%%%%%%%%%%%%%%%%%%%%%%%%%%%%%%%%%%%%%%%%%%%%%%%%%%%%%%%%%%%%%%%%%%%%%

\smallskip
We start our phenomenological analysis with the preferred parameter
spaces found in our scan in the \sihi scenario. 
In the description of the plots we will focus on the red points, i.e.\
the ones that potentially describe the excesses observed at ATLAS and
CMS. Only if the other levels of exclusion exhibit important
information we will comment on them.

\begin{figure}[htb!]
\begin{subfigure}[b]{\linewidth}
\centering\includegraphics[width=0.60\textwidth]{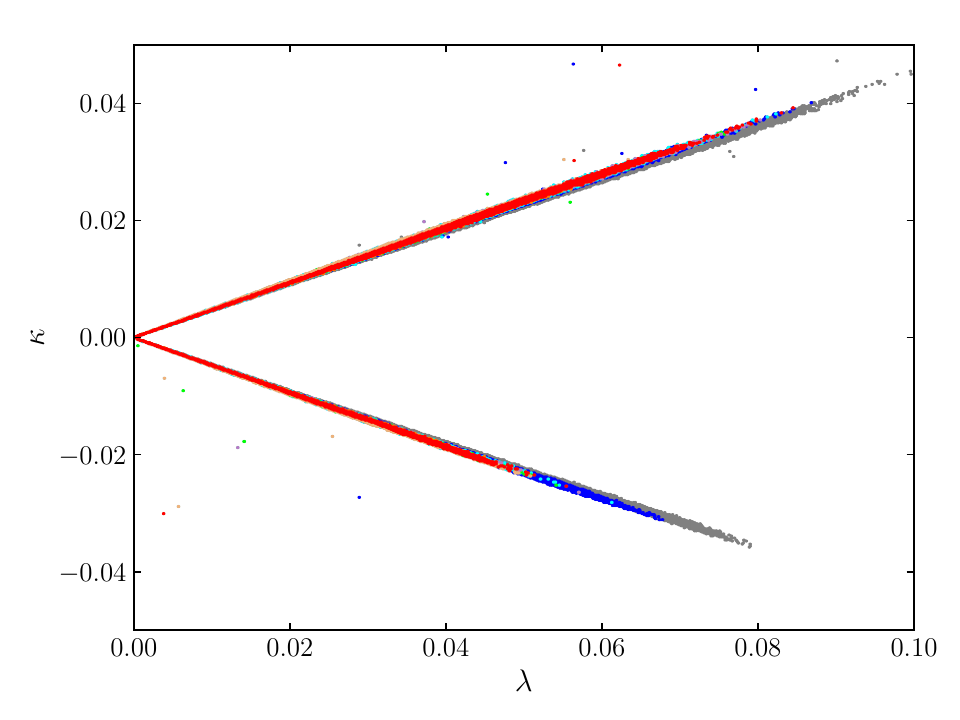}
\end{subfigure}\\
\begin{subfigure}[b]{\linewidth}
\centering\includegraphics[width=\textwidth]{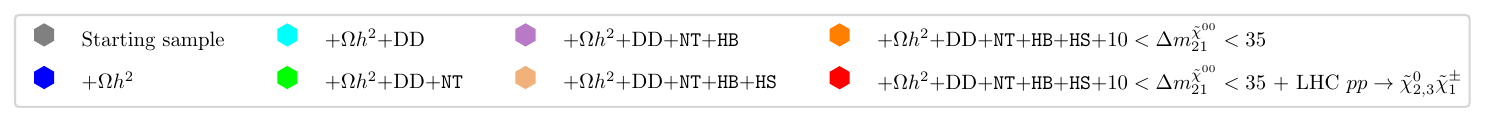}
\end{subfigure}      
\caption{The results of our parameter scan in the \sihi scenario
in the $\la$--$\ka$ plane.}
\label{fig:scan}
\end{figure}

In \reffi{fig:scan} we show the coverage of our parameter scan in
the \sihi scenario for 
the $\la$--$\ka$ plane. 
The plot demonstrates that $\la$ and $\ka$ are forced into a narrow range of
small values with a strong correlation among these two
parameters. This is driven by the requirement to respect
the DM relic density constraint, which selects values for which $\lambda \simeq 2 \kappa$ (see, e.g., \citere{Agin:2024yfs}).
Concerning $A_\la$ (not shown), it covers the whole scan range, whereas $A_\ka$ (not shown) is restricted to
$|A_\ka| \lsim 350 \gev$. 

%%%%%%%%%%%%%%%%%%%%%%%%%%%% F I G U R E %%%%%%%%%%%%%%%%%%%%%%%%%%%%%%
%%%%%%%%%%%%%%%%%%%%%%%%%%%% F I G U R E %%%%%%%%%%%%%%%%%%%%%%%%%%%%%%

%%%%%%%%%%%%%%%%%%%%%%%%%%%% F I G U R E %%%%%%%%%%%%%%%%%%%%%%%%%%%%%%
\begin{figure}[htb!]
\centering
\centering\includegraphics[width=0.8\textwidth]{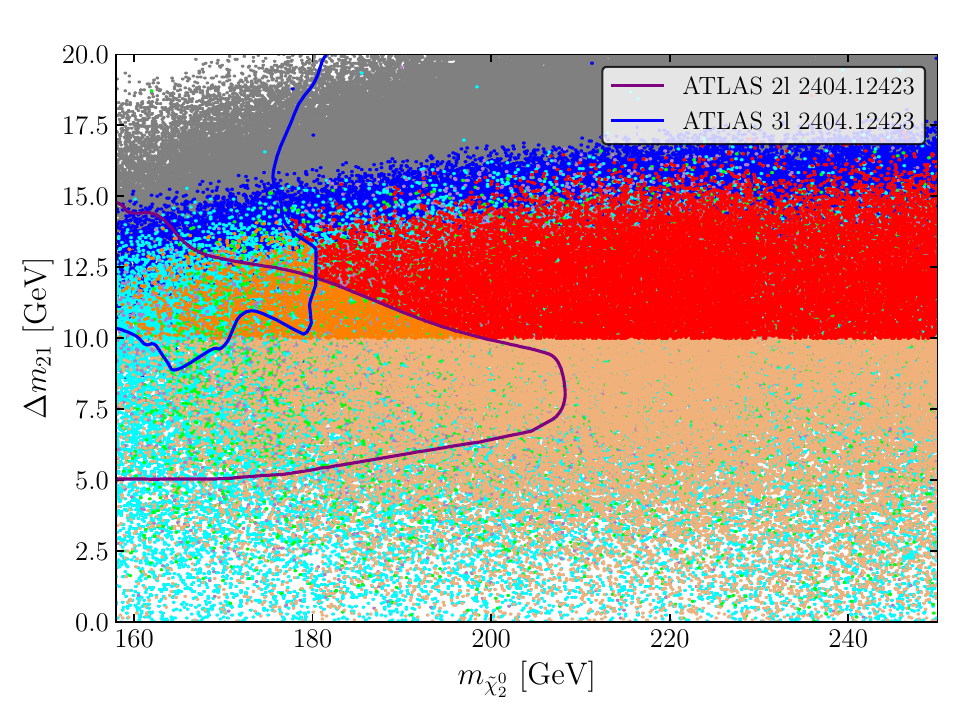}\\
\centering\includegraphics[width=\textwidth]{legend.pdf}\\
\caption{The results of our parameter scan in the \sihi scenario
in the $\mneu2$--$\De m_{21}$ plane.}
\label{fig:mneu2-Dem}
\end{figure}
%%%%%%%%%%%%%%%%%%%%%%%%%%%% F I G U R E %%%%%%%%%%%%%%%%%%%%%%%%%%%%%%

In \reffi{fig:mneu2-Dem} we show the main result of our parameter
scan: the $\mneu2$--$\De m_{21}$ plane, overlaid with the experimental 95\%
CL exclusion 
bounds as recasted in \citere{Agin:2024yfs}. The blue line indicates
the ATLAS 3~soft-lepton limits~\cite{ATLAS:2021moa}, whereas the purple
line indicates the ATLAS 2~soft-lepton limits~\cite{ATLAS:2019lng}.%
\footnote{The recasting performed in \citere{Agin:2024yfs} was performed only for the ATLAS limits.
However, it has been discussed in the literature, see e.g.\ \citere{CHS6}, that the ATLAS and CMS limits
effectively show the same excesses in the 3~and 2~soft-lepton channels.}
~One can observe that
our red points, i.e.\ the ones passing all constraints stretch out
from the 95\% CL exclusion lines up to $\mneu2 = 250 \gev$, where our
scan stopped. $\De m_{21} := \mneu2 - \mneu1$ is found between $10 \gev$
(where we cut the ``preferred parameter space'') 
and $\sim 16 \gev$, with the values closest to the exclusion bounds of
$12 \gev \lsim \De m_{21} \lsim 14 \gev$. This plot clearly demonstrates
that the \sihi scenario in the NMSSM yields a very good description of
the soft-lepton excesses observed at ATLAS and CMS, while being in
agreement with all other theoretical and experimental constraints.

\begin{figure}[htb!]
%       \vspace{1em}
\centering
\begin{subfigure}[b]{0.48\linewidth}
\centering\includegraphics[width=\textwidth]{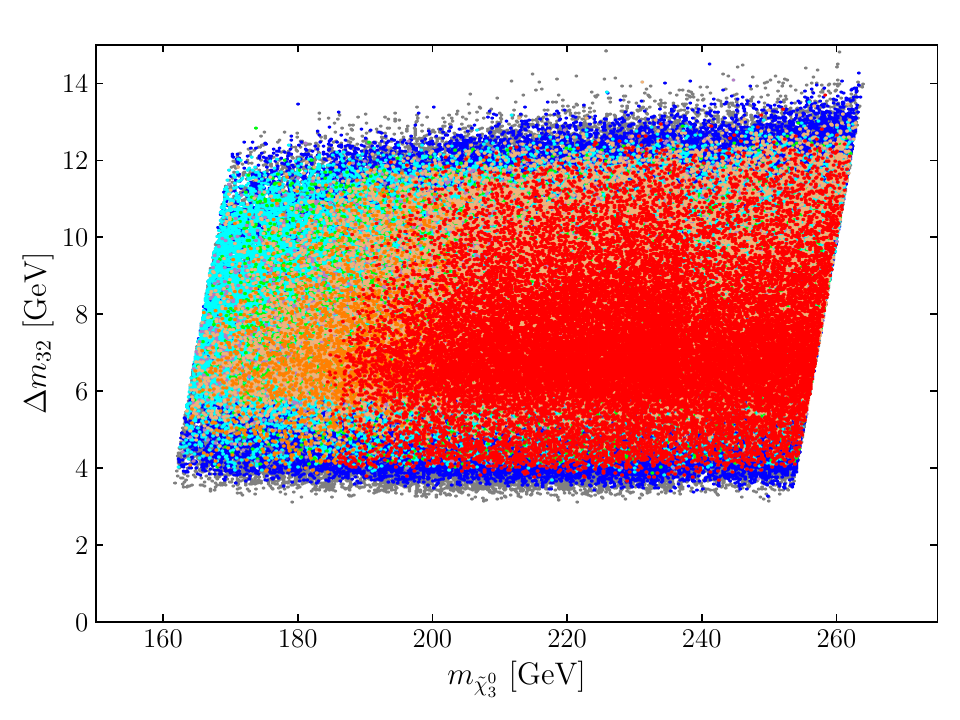}
\end{subfigure}
~
\begin{subfigure}[b]{0.48\linewidth}
\centering\includegraphics[width=\textwidth]{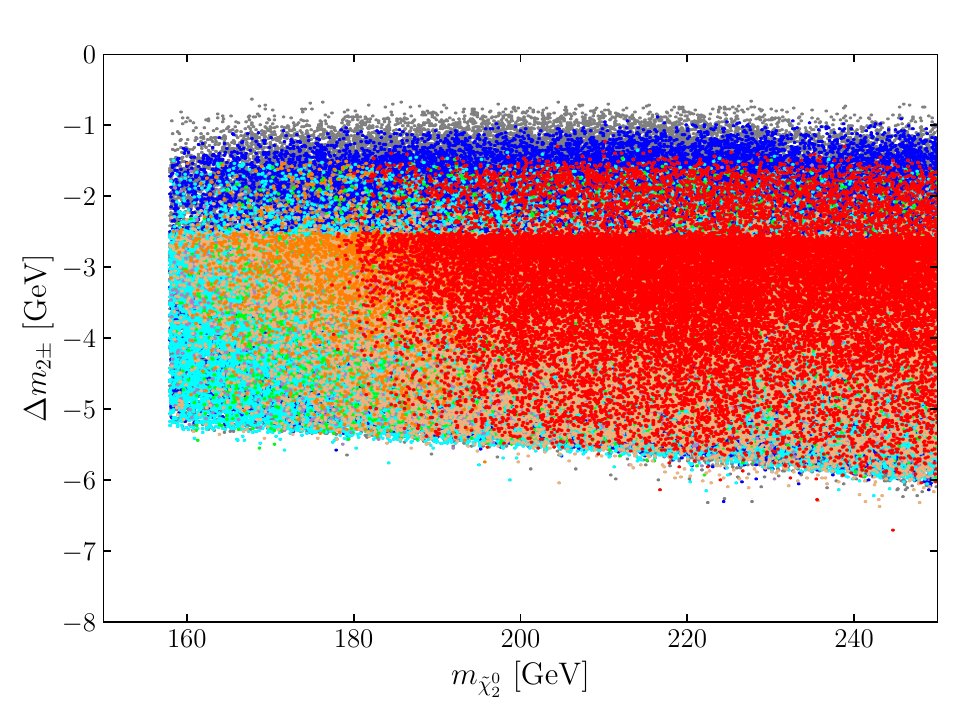}
\end{subfigure}\\
\begin{subfigure}[b]{\linewidth}
\centering\includegraphics[width=\textwidth]{legend.pdf}
\end{subfigure}
\caption{The results of our parameter scan in the \sihi scenario
in the $\mneu2$--$(\mneu3-\mneu2)$ plane (left) and the
$\mneu2$--$(\mneu2-\mcha1)$ plane (right).}
\label{fig:otherDm}
\end{figure}

The fact that $\De m_{21}$ comes out somewhat smaller in the NMSSM as
compared to the preferred $\sim 20 \gev$ in the MSSM (see,
e.g., \citere{CHS6}) can be attributed to the fact that two production
channels are open in the NMSSM, $pp \to \neu{2,3}\cha1$, as compared
to only one in the MSSM, where by definition we have $\mneu3 > \mneu2$.
This is quantified in the left plot of \reffi{fig:otherDm}, where we
show the results of our scan in the $\mneu2$--$(\mneu3-\mneu2)$ plane. 
One can see that the preferred parameter space has
$3 \gev \lsim \mneu3 - \mneu2 \lsim 13 \gev$, leading to a
correspondingly larger difference in $\mneu3 - \mneu1$.
In the right plot of \reffi{fig:otherDm} we show our scan results in
the $\mneu2$--$(\mneu2 - \mcha1)$ plane. Here one can observe that the
lightest chargino is up to $\sim 5 \gev$ heavier than the second
lightest neutralino.

%%%%%%%%%%%%%%%%%%%%%%%%%%%% F I G U R E %%%%%%%%%%%%%%%%%%%%%%%%%%%%%%
%%%%%%%%%%%%%%%%%%%%%%%%%%%% F I G U R E %%%%%%%%%%%%%%%%%%%%%%%%%%%%%%

\begin{figure}[htb!]
\centering
\centering\includegraphics[width=0.6\textwidth]{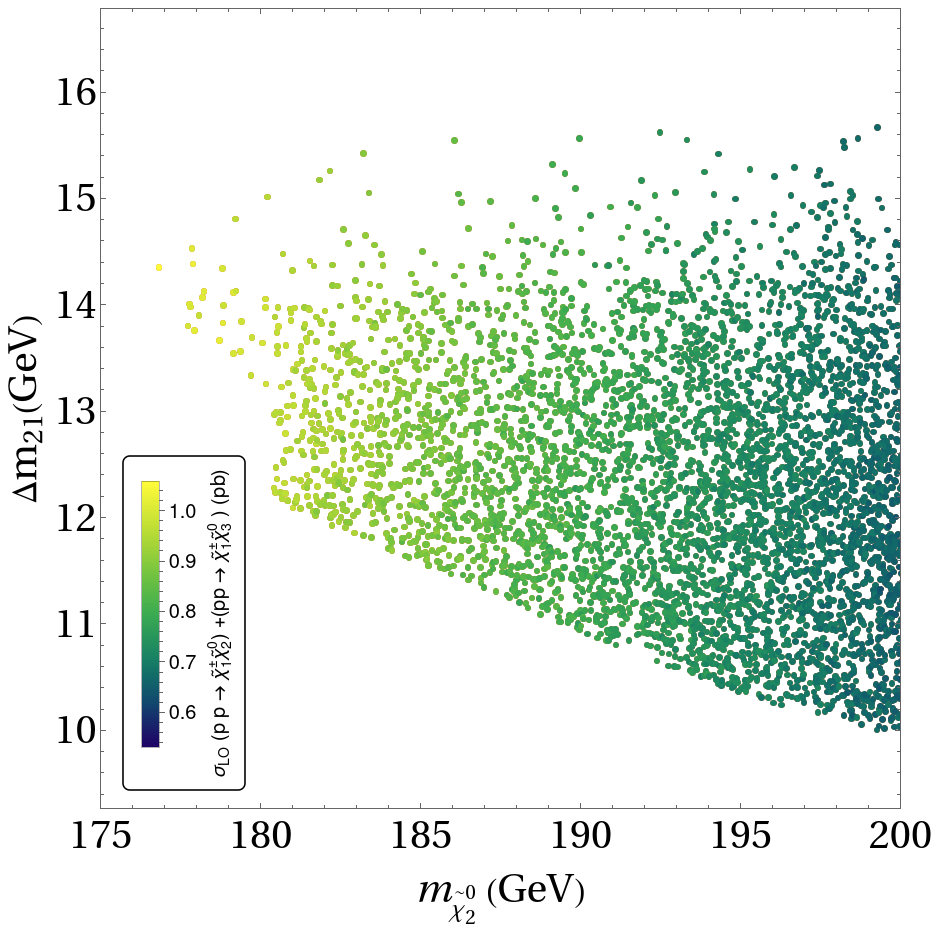}
\caption{The predictions of $\sig(pp \to \neu2\cha1) + \sig(pp \to \neu3\cha1)$
(see text) are shown as color coding in the $\mneu2$-$\De m_{21}$ plane.}
\label{fig:mneu2-si23}
\end{figure}

In \reffi{fig:mneu2-si23} we show our predictions for
$\sig(pp \to \neu2\cha1) + \sig(pp \to \neu3\cha1)$ at the LHC at
$\sqrt{s} = 13 \tev$ in the $\mneu2$--$\De m_{21}$ plane.
The combined cross-section underscores that contributions from both Higgsino-like NLSPs, 
$\neu2$ and $\neu3$ are relevant for the signal channel.
We computed the leading-order (LO) production cross-section using {\tt MadGraph (MG5aMC)-v3.6.3}~\cite{Alwall:2014hca} for which we 
incorporated the NMSSM model file from the {\tt FeynRules}~\cite{Alloul:2013bka,Conte:2016zjp} database. 
The SLHA files, generated from our {\tt NMSSMTools} run and validated against all relevant constraints, 
were passed as {\tt MadGraph} param-cards. To obtain the NLO cross-section, we applied the MSSM k-factor given by 
{\tt Resummino}~\cite{resummino,Bozzi:2006fw,Bozzi:2007qr,Debove:2009ia,Debove:2010kf} to the LO production cross-section. 
We assume the QCD loop effects in both the NMSSM and MSSM production cross-sections to be of similar order.
The size of the cross
section for each point is indicated by the color scale. One can see that
the cross section mainly depends on $\mneu2 \approx \mcha1$, but is
nearly independent on $\De m_{21}$:
while an increasing mass of the final state particles leads to smaller cross sections, $\De m_{21}$ has only
a small impact on the mixings of $\neu2$ and thus its couplings entering the cross section calculation.
The smallest EWino masses yield production cross sections of approximately $\sim 1.3 \pb$, going down to $\sim 0.7 \pb$
for $\mneu2 \sim 200 \gev$ (where our plot ends).
These values are roughly consistent with the cross sections required to reproduce the observed excess in events, within the associated 
uncertainties, see e.g.\ the discussion in \citere{CHS6}.
On the other hand, this defines a clear target for Run~3 of the LHC and future collider stages. Although no dedicated 
projections for the discovery potential at the HL-LHC are currently available, it appears plausible that Run~3 could provide valuable insights 
into the proposed explanation of the soft-lepton excesses.

%%%%%%%%%%%%%%%%%%%%%%%%%%%% F I G U R E %%%%%%%%%%%%%%%%%%%%%%%%%%%%%%
%%%%%%%%%%%%%%%%%%%%%%%%%%%% F I G U R E %%%%%%%%%%%%%%%%%%%%%%%%%%%%%%

\begin{figure}[htb!]
\centering
\begin{subfigure}[b]{0.48\linewidth}
\centering\includegraphics[width=\textwidth]{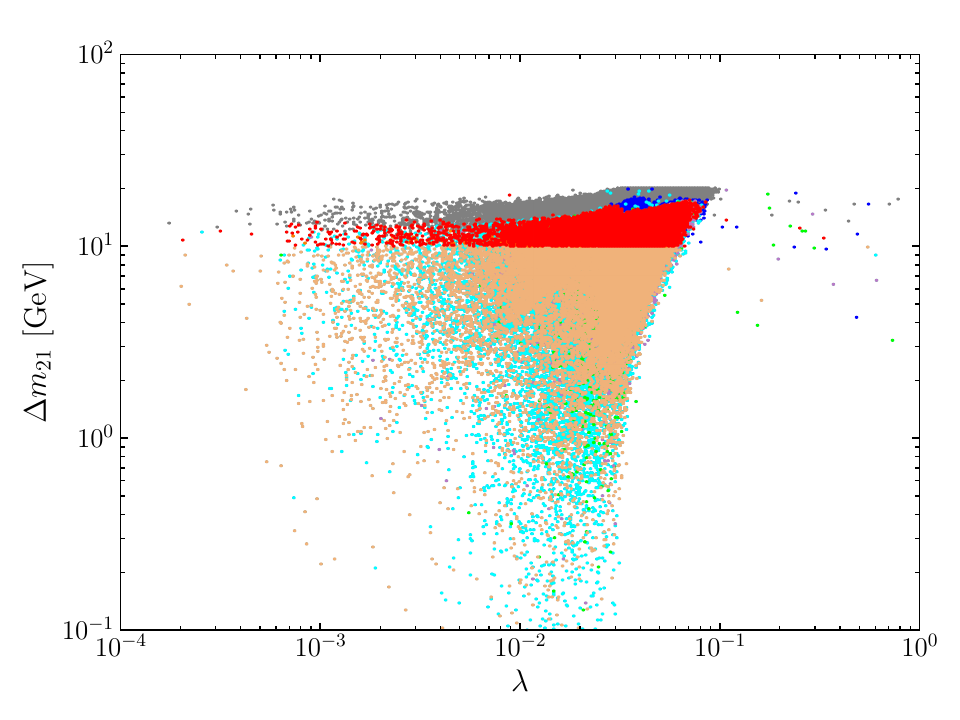}
\end{subfigure}
~
\begin{subfigure}[b]{0.48\linewidth}
\centering\includegraphics[width=\textwidth]{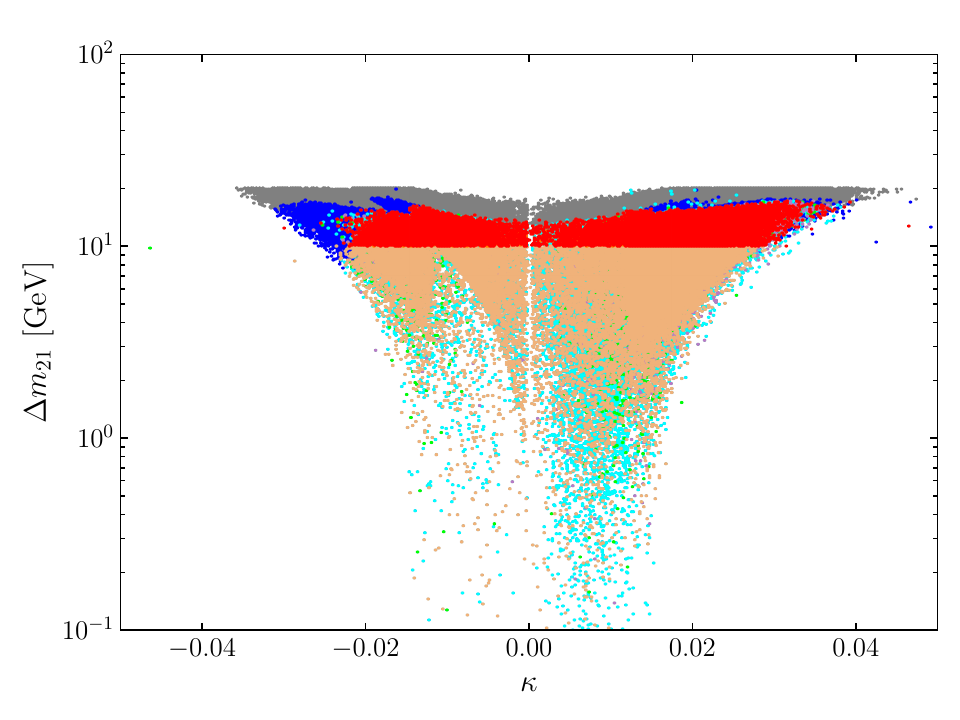}
\end{subfigure}\\
\begin{subfigure}[b]{\linewidth}
\centering\includegraphics[width=\textwidth]{legend.pdf}
\end{subfigure}      
\caption{The results of our parameter scan in the \sihi scenario
in the $\la$--$\De m_{21}$ plane (left) and the $\ka$--$\De m_{21}$ plane
(right). }
\label{fig:la-ka-Dm}
\end{figure}

In \reffi{fig:la-ka-Dm}, we present our results in the $\lambda$–$\De m_{21}$ plane (left) and the $\kappa$–$\De m_{21}$ plane (right). It can be seen that larger values of $\De m_{21}$ are attained for higher $\lambda$, although the dependence remains relatively mild. The preferred region corresponds to $0.03 \lesssim \lambda \lesssim 0.08$. In contrast, no clear trend is observed for $\kappa$, apart from a dip in the (red) allowed parameter space that excludes $\kappa$ values near zero for the largest $\De m_{21}$.

%%%%%%%%%%%%%%%%%%%%%%%%%%%% F I G U R E %%%%%%%%%%%%%%%%%%%%%%%%%%%%%%
%%%%%%%%%%%%%%%%%%%%%%%%%%%% F I G U R E %%%%%%%%%%%%%%%%%%%%%%%%%%%%%%

%%%%%%%%%%%%%%%%%%%%%%%%%%%% F I G U R E %%%%%%%%%%%%%%%%%%%%%%%%%%%%%%
\begin{figure}[htb!]
\centering
\begin{subfigure}[b]{0.48\linewidth}
\centering\includegraphics[width=\textwidth]{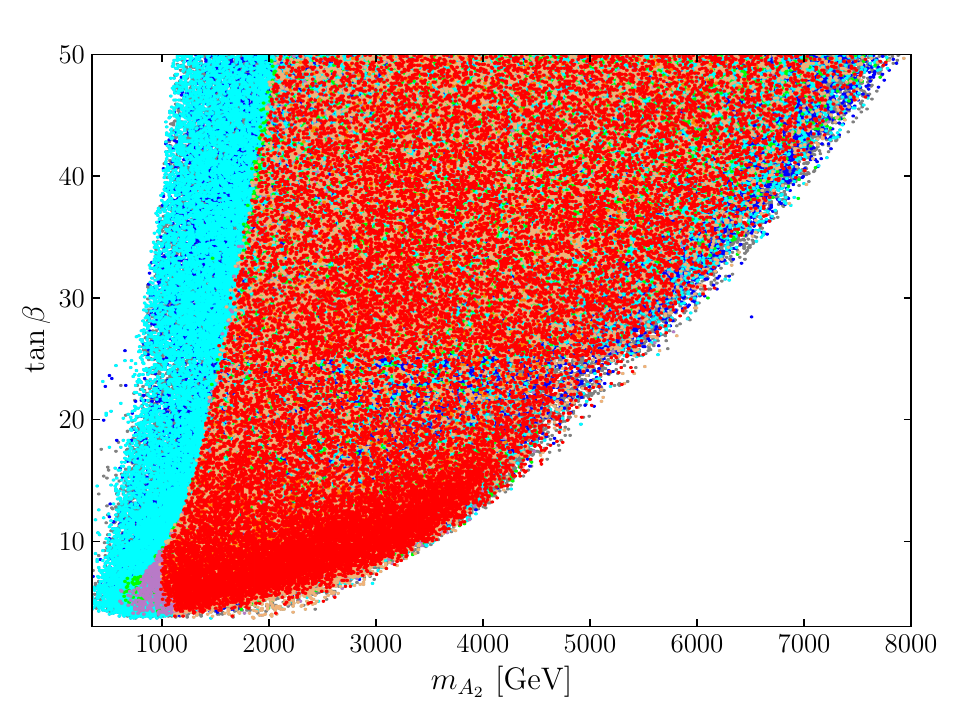}
\end{subfigure}
~
\begin{subfigure}[b]{0.48\linewidth}
\centering\includegraphics[width=\textwidth]{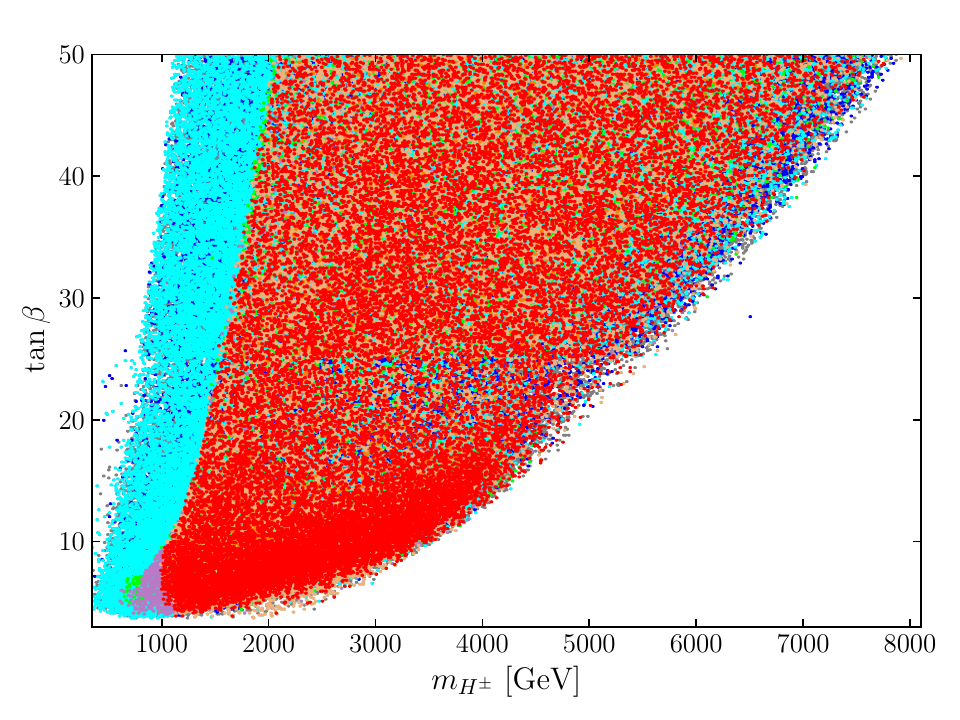}
\end{subfigure}\\
\begin{subfigure}[b]{\linewidth}
\centering\includegraphics[width=\textwidth]{legend.pdf}
\end{subfigure}      
\caption{The results of our parameter scan in the \sihi scenario
in the $\mAz$--$\tb$ plane (left) and the $\mHp$--$\tb$ plane
(right). }
\label{fig:ma-tb}
\end{figure}
%%%%%%%%%%%%%%%%%%%%%%%%%%%% F I G U R E %%%%%%%%%%%%%%%%%%%%%%%%%%%%%%

We now turn to the results for other particle masses in our scan. As
discussed above, the scalar leptons and quarks are chosen to be heavy,
and we do not present any results for them here. On the other hand,
the heavy NMSSM Higgs-boson masses are effectively treated as free
parameters, subject to the BSM Higgs searches as implemented
into \HB\ as well as the LHC Higgs-boson rate measurements as
implemented into \HS. In \reffi{fig:ma-tb} we show the results of our
scan in the 
$\mAz$--$\tb$ plane (left)  and the $\mHp$--$\tb$ plane (right).
In the left plot one can see that $\mAz$ ranges from $\sim 1000 \gev$
up to values $\gsim 7.5 \tev$ for $\tb = 50$, where our scan stopped.
However, one can clearly
observe the parameter space that is excluded by the BSM Higgs-boson
searches, starting at $\mAz \sim 1000 \gev$ and $\tb \sim 10$. A
small region of purple points is visible at low $\tb$ for
$\mAz \lsim 1000 \gev$, where points are excluded by \HS. The large
cyan region visible for lower $\mAz$ values and $\tb \gsim 10$ are
excluded by the searches for $pp \to \phi \to \tau\tau$, with $\phi$
being a heavy $\cp$-even or $\cp$-odd Higgs boson. 
Apart from this, no specific restriction on the heavy $\cp$-odd Higgs-boson mass can be identified, implying that 
no clear prediction for future collider searches can be made. The charged Higgs-boson mass, shown in the right panel 
of the figure, is strongly correlated with the neutral Higgs-boson masses and therefore exhibits a similar behavior. 
However, the direct limits on charged Higgs bosons are less stringent than those on neutral Higgs bosons, and the 
observed constraints on the charged Higgs sector mainly originate from searches targeting the neutral Higgs bosons.

In \reffi{fig:mH-single} we turn to the lighter $\cp$-even masses in
our scan. In the left plot we show the results in the
$\mHe$--$|S_{13}|^2$ plane, i.e.\ the light $\cp$-even Higgs-boson
mass together with its singlet component. We find
$122 \gev \lsim \mHe \lsim 128 \gev$, as required by the LHC
Higgs-boson measurements (and here enforced by, taking into account a $3 \gev$ theory uncertainty~\cite{Degrassi:2002fi}). 
The singlet component is always smaller than 10\% (with even smaller values for
$\mHe$ close to the allowed lower and upper limits). Also these small
values are imposed by \HS\ and ensure the agreement with the LHC
measurements. The right plot shows the $\mHz$--$|S_{23}|^2$ plane,
i.e.\ the second lightest $\cp$-even Higgs-boson mass together with
its singlet component. We find that the mass is restricted to be
$125 \gev \lsim \mHz \lsim 225 \gev$, with a singlet component of 90\%
or larger. This indicates that the soft-lepton excesses at ATLAS and
CMS, when interpreted in the NMSSM, predict a second light
Higgs-boson, which, however, largely decouples from the SM
particles. {Nevertheless, this light Higgs-boson might be an interesting
target for future HL-LHC direct searches. We leave a detailed
phenomenological analysis for future work.

%%%%%%%%%%%%%%%%%%%%%%%%%%%% F I G U R E %%%%%%%%%%%%%%%%%%%%%%%%%%%%%%
\begin{figure}[htb!]
\centering
\begin{subfigure}[b]{0.48\linewidth}
\centering\includegraphics[width=\textwidth]{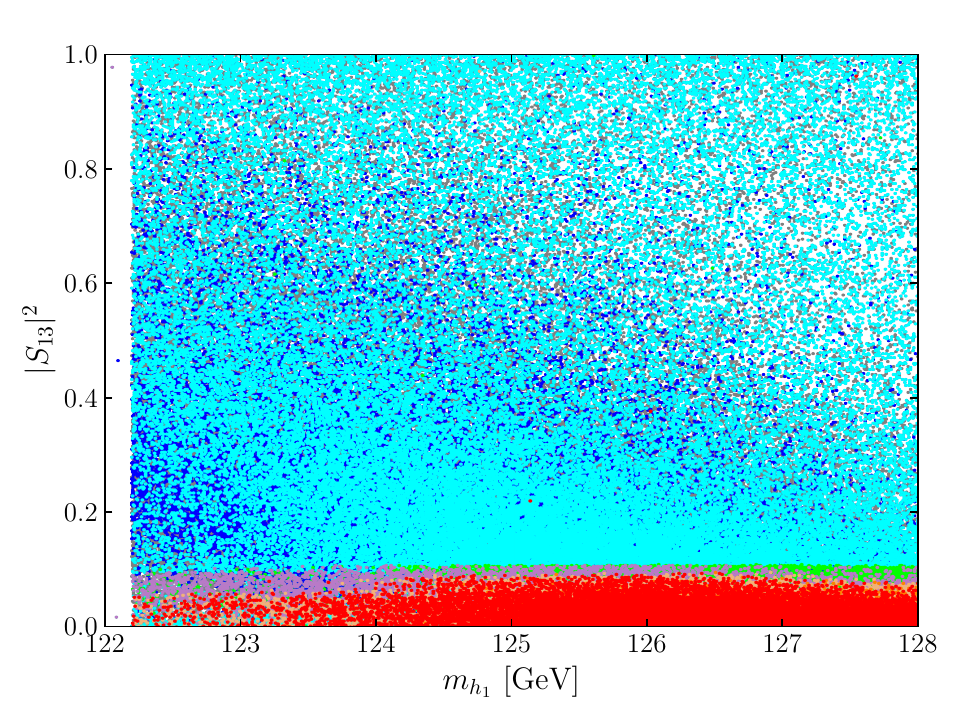}
\end{subfigure}
~
\begin{subfigure}[b]{0.48\linewidth}
\centering\includegraphics[width=\textwidth]{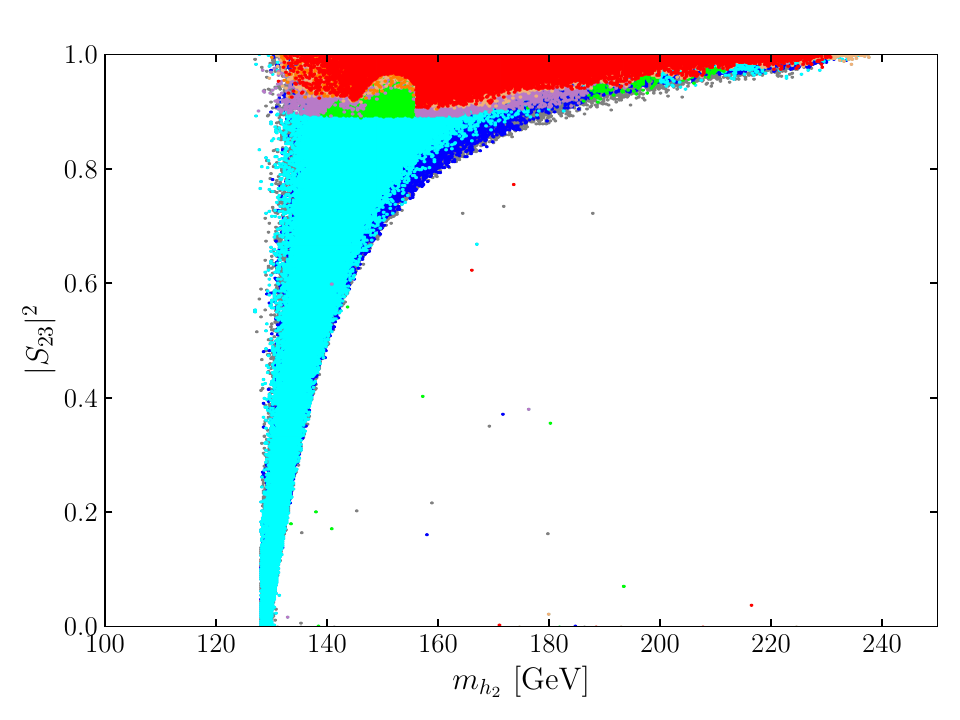}
\end{subfigure}\\
\begin{subfigure}[b]{\linewidth}
\centering\includegraphics[width=\textwidth]{legend.pdf}
\end{subfigure}      
\caption{The results of our parameter scan in the \sihi scenario
in the $\mHe$--$|S_{13}|^2$ plane (left) and the
$\mHz$--$|S_{23}|^2$ plane (right).}
\label{fig:mH-single}
\end{figure}
%%%%%%%%%%%%%%%%%%%%%%%%%%%% F I G U R E %%%%%%%%%%%%%%%%%%%%%%%%%%%%%%

We finish our analysis of the $\cp$-even Higgs bosons
with \reffi{fig:mcha1-gaga}, where we show $g_{\ga\ga}$ as a function
of $\mcha1$. $g_{\ga\ga}$ is the effective coupling of the $h_1$ to two photons,
and it is affected by the contributions from light charged particles that would appear in loop diagrams in the full theory. 
The lightest chargino naturally can
contribute to a deviation of $g_{\ga\ga}$ from unity. However, one can
observe that this deviation hardly exceeds the 5\% level, and the
predicted rate of $\He \to \ga\ga$ is close to the SM prediction, and
thus in agreement with the experimental data. However, part of this parameter space
which features relatively small deviations will be probed by HL-LHC and future colliders. On the other hand, 
it is apparent that parameter points indistinguishable from the SM at any foressen experiment are also possible. 

%%%%%%%%%%%%%%%%%%%%%%%%%%%% F I G U R E %%%%%%%%%%%%%%%%%%%%%%%%%%%%%%
\begin{figure}[htb!]
\centering
\centering\includegraphics[width=0.6\textwidth]{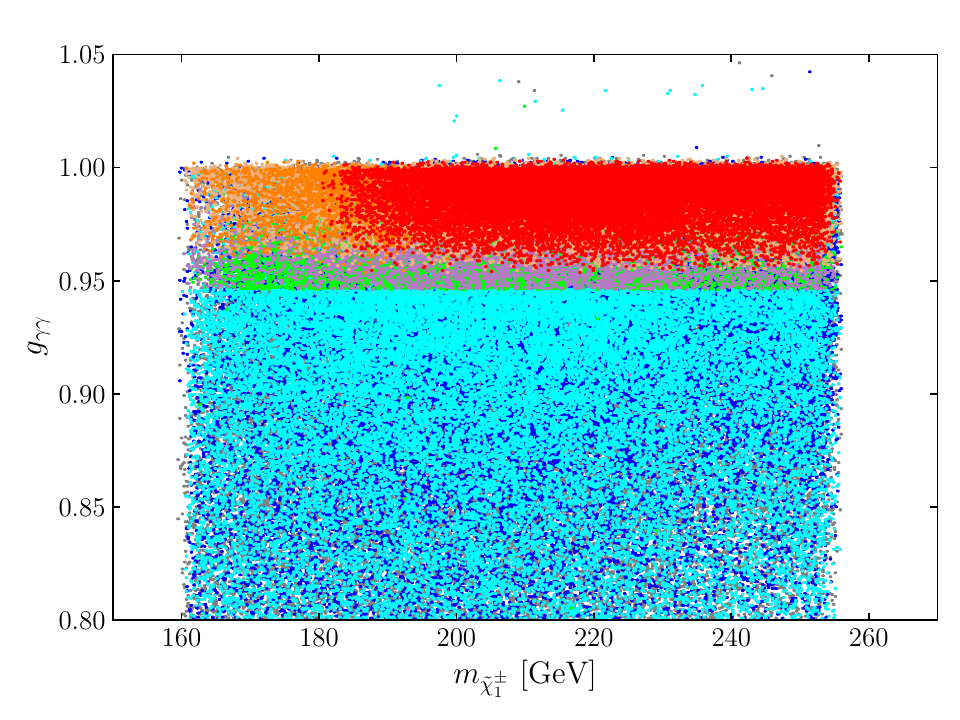}\
\centering\includegraphics[width=\textwidth]{legend.pdf}\\
\caption{The results of our parameter scan in the \sihi scenario
in the $\mcha1$--$g_{\ga\ga}$ plane.}
\label{fig:mcha1-gaga}
\end{figure}
%%%%%%%%%%%%%%%%%%%%%%%%%%%% F I G U R E %%%%%%%%%%%%%%%%%%%%%%%%%%%%%%

%%%%%%%%%%%%%%%%%%%%%%%%%%%% F I G U R E %%%%%%%%%%%%%%%%%%%%%%%%%%%%%%
\begin{figure}[htb!]
\centering
\centering\includegraphics[width=0.80\textwidth]{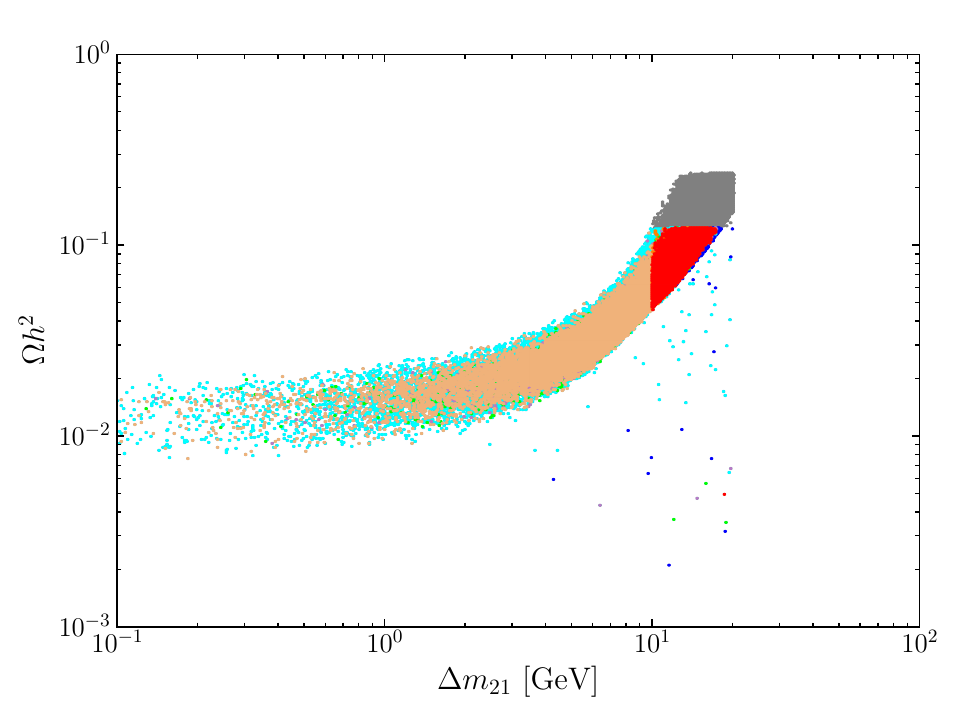}\\
\centering\includegraphics[width=\textwidth]{legend.pdf}\\
\caption{The results of our parameter scan in the \sihi scenario
in the $\De m_{21}$--$\Och$ plane.}
\label{fig:Dem-Oh2}
\end{figure}
%%%%%%%%%%%%%%%%%%%%%%%%%%%% F I G U R E %%%%%%%%%%%%%%%%%%%%%%%%%%%%%%

Now we turn to the predictions for DM in our \sihi
scenario. In \reffi{fig:Dem-Oh2} we present our scan results in
the $\De m_{21}$--$\Och$ plane. A clear correlation over the whole
scanned parameter space is visible, with lower DM relic densities reached for
smaller $\De m_{21}$. Turning this around, larger $\De m_{21}$, as favored by
the soft-lepton excesses leads naturally to a (possible) saturation of
the DM relic density. This is in stark contrast to the higgsino
scenarios in the MSSM, see \citere{CHS6}, where always a strongly
underabundant DM relic density was found.

%%%%%%%%%%%%%%%%%%%%%%%%%%%% F I G U R E %%%%%%%%%%%%%%%%%%%%%%%%%%%%%%
\begin{figure}[htb!]
\centering
\centering\includegraphics[width=0.6\textwidth]{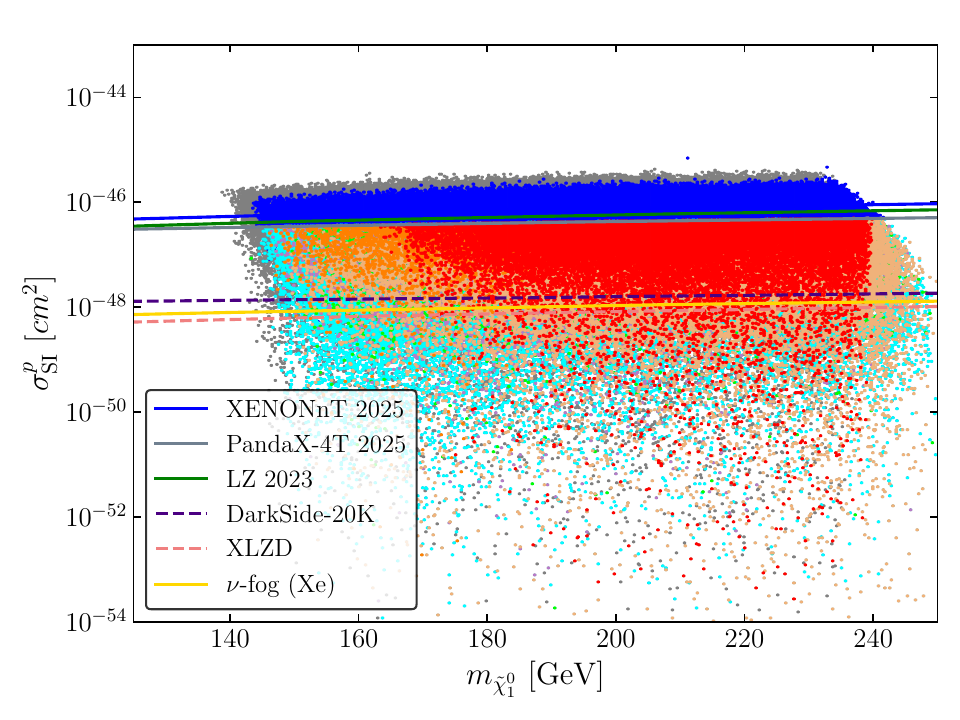}
\centering\includegraphics[width=0.48\textwidth]{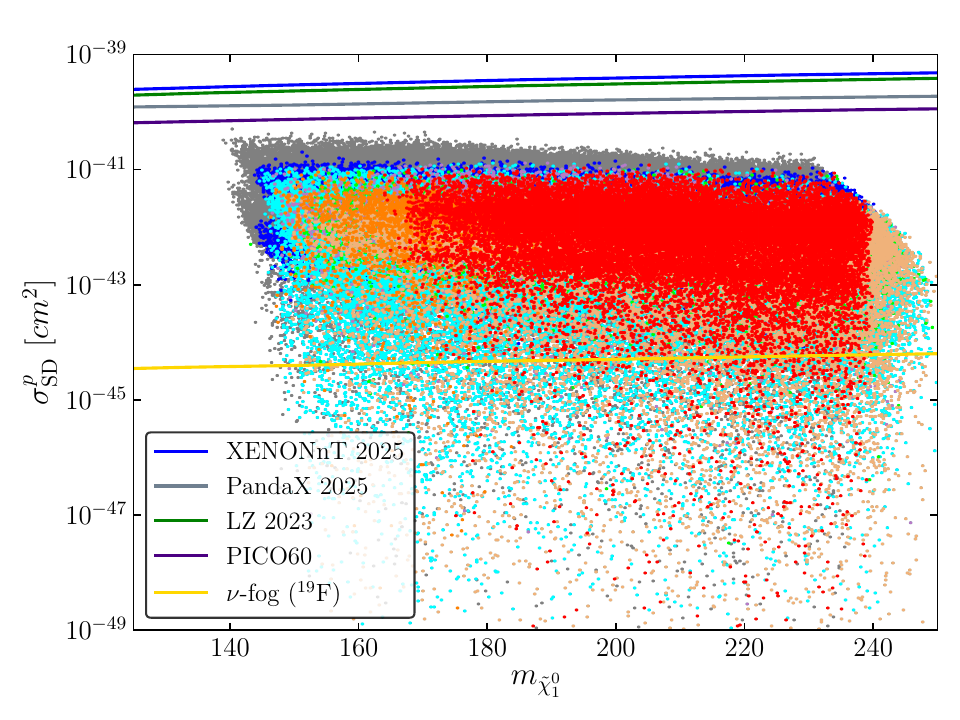}
          \includegraphics[width=0.48\textwidth]{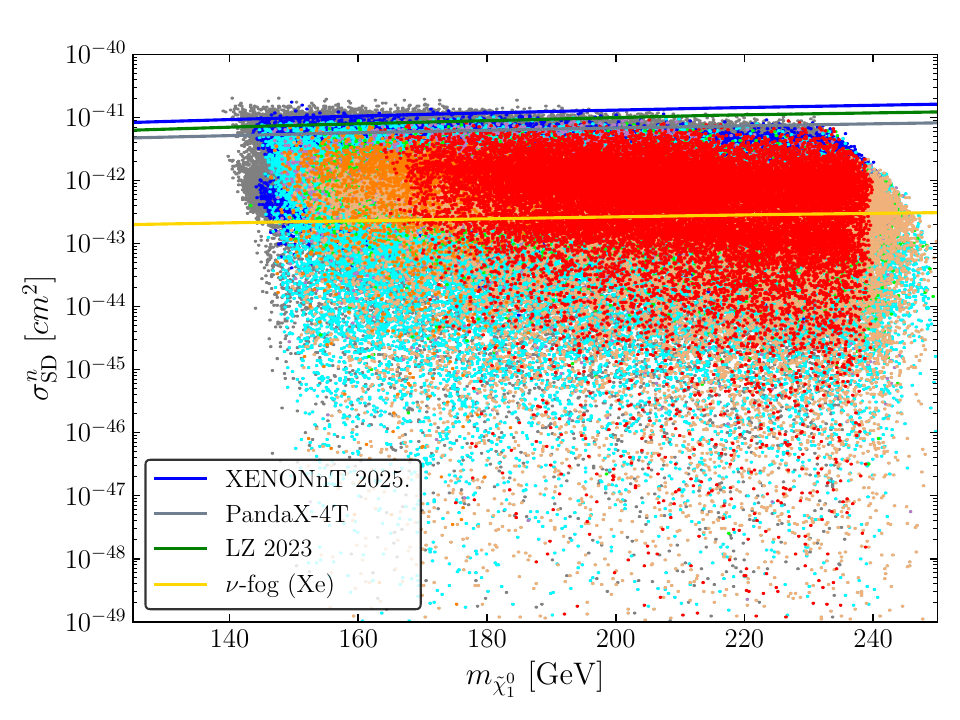}\\
\centering\includegraphics[width=\textwidth]{legend.pdf}\\
\caption{The results of our parameter scan in the \sihi scenario
in the $\mneu1$--$\ssi$ plane (upper plot) and the $\mneu1$--$\ssdp$ plane
(lower left plot) and the $\mneu1$--$\ssdn$ plane (lower right plot)}.
\label{fig:mneu1-ssi-ssd}
\end{figure}
%%%%%%%%%%%%%%%%%%%%%%%%%%%% F I G U R E %%%%%%%%%%%%%%%%%%%%%%%%%%%%%%

The prospects for spin-independent ($\ssi$) and spin-dependent 
($\ssdp$, $\ssdn$ for $p$ and $n$ scattering, respectively) direct DM
searches in our \sihi scenario are presented in \reffi{fig:mneu1-ssi-ssd}. 
All plots show the respective cross section
as a function of the DM mass, $\mneu1$.
In the upper plot we show $\ssi$ and indicate with solid blue, black and gray lines
the current bounds from XENON-nT~\cite{XENON:2025vwd}, PandaX-4T~\cite{PandaX:2024qfu} and LZ~\cite{LZ:2024zvo}, 
respectively. As discussed in \refse{sec:constraints} and visible in the plot, we
applied the limits from PandaX-4T~\cite{PandaX:2024qfu}. Having applied the more recent published limits 
from LZ~\cite{LZ:2024zvo} would cut away the points with the highest DD cross sections. However, we do not
expect a major relevant impact on our general findings. 
As dashed purple and red lines we indicate the anticipated limits from DarkSide-20K~\cite{DarkSide-20k:2017zyg} and
XLZD~\cite{XLZD:2024nsu}, respectively. The $\nu$-fog (for Xenon)~\cite{Billard:2013qya,Ruppin:2014bra} is shown as yellow solid line.
While the future experiments can cover a large part of the preferred parameter space, the NMSSM
interpretation of the 2- and 3-soft lepton excesses are also compatible with points predicting a DD 
cross section below the $\nu$-fog limit, i.e.\ not observable with currently used experimental set-ups (see also \citere{Bhattiprolu:2025zwt} for a recent analysis of higgsino DM in DD experiments).
We have tested that the points below the $\nu$-fog limit correspond to somewhat smaller $\De m_{21}$, not 
exceeding $13 \gev$ for the smallest allowed $\mneu2$ values.
In the lower row of \reffi{fig:mneu1-ssi-ssd} on the left (right) we present the results for the 
spin-dependent scattering cross section with protons (neutrons). The blue, gray and green solid lines indicate the
limits from XENON-nT~\cite{XENON:2025vwd}, PandaX-4T~\cite{PandaX:2024qfu} and LZ~\cite{LZ:2024zvo}, 
respectively. In the left plot we indicate additionally the limit from PICO60~\cite{PICO:2019vsc} 
as solid black line. 
The yellow solid lines show the $\nu$-fog limits for $^{19}$F (left) and Xenon (right)~\cite{Billard:2013qya,Ruppin:2014bra}.
One can observe that the SD cross section with protons is predicted to be far below any current bounds, and can also reach 
below the $^{19}$F $\nu$-fog limit. On the other hand, the SD scattering with neutrons reaches the strongest experimental 
limit, as currently given by PandaX-4T, which cuts away a couple of points of our scan. Also this cross section reaches well 
below the Xenon $\nu$-fog limit. Overall, while the DD experiments show interesting prospects for the future runs, 
the NMSSM singlino/higgsino explanation of the soft lepton excesses may well escape this kind of experimental test 
(in contrast to the MSSM explanations presented in \citere{CHS6}).

%%%%%%%%%%%%%%%%%%%%%%%%%%%%%%%%%%%%%%%%%%%%%%%%%%%%%%%%%%%%%%%%%%%%%%%%%%
%%%%%%%%%%%%%%%%%%%%%%%%%%%%%%%%%%%%%%%%%%%%%%%%%%%%%%%%%%%%%%%%%%%%%%%%%%

\section{Conclusions}
\label{sec:conclusion}

The EW sector of the NMSSM, consisting of five neutralinos,
$\neu{1,2,3,4,5}$ and two charginos, $\cha{1,2}$ (as well as the SUSY partners
of the SM leptons), 
can account for a variety of experimental data that cannot be
described by the SM. In particular, assuming that the lightest
neutralino, $\neu1$, is the LSP, this particle constitutes a DM candidate
in good agreement with the observed limits on the DM relic density of the
universe, as well as with negative results from direct detection experiments. 

At the LHC, a wide range of experimental searches are carried out
to search for the production and decays of the EWinos in various final
states, often performed in the context of the MSSM.
In particular, the ``golden channel'', 
$pp \to \neu{2},\cha1 \to \neu1 Z^{(*)} \, \neu1 W^{\pm(*)}$ is used
to place stringent constraints on the EW MSSM parameter space
(where sleptons are assumed to be heavy).
The targeted final states contain two or three (possibly soft) leptons
accompanied by a substantial amount of $\ETmiss$. 
These searches are particularly challenging in the part of the
parameter space where the mass differences 
between the final and the initial state EWinos become small, leading
to rather soft final state leptons.
Despite these experimental challenges, the
CMS and ATLAS collaborations are searching actively for 
EWinos in this ``compressed spectra'' region, focusing on the 2~and
3~soft-lepton plus $\ETmiss$ final state.
Interestingly, the experimental data of searches in the ``golden channel'',
$pp \to \neu2 \cha1 \to \neu1 Z^{*} \, \neu1 W^{\pm *}$ show consistent
excesses between CMS and ATLAS in exactly these final
states~\cite{ATLAS:2019lng,ATLAS:2021moa,CMS:2024gyw,CMS:2025ttk}. 
These searches assume a simplified model scenario with
$\mneu2 \approx \mcha1$ and
$\De m_{21} := \mneu2 - \mneu1 \approx \order{20 \gev}$.
The sleptons are assumed to be relatively heavy, and thus to not take
part in the decays of the initially produced EWinos. 

In \citere{CHS6} within the MSSM three different scenarios were analyzed,
classified by the hierarchy and signs of the EWino mass parametes,
$M_1$, $M_2$ and $\mu$, while assuming heavy sleptons.
It was found that the MSSM scenarios that give a good
description of the soft-lepton excesses and complying with the DM
experimental constraints, require $|M_1| \sim |M_2|$. 
However, one of the most compelling arguments in favor of SUSY is the
unification of forces at the GUT scale, yielding (in simple GUT
realization) the following mass pattern at the EW
scale: $M_1 \sim M_2/2 \sim M_3/6$, where $M_3$ is the gluino mass, $\mgl$.
Searches for colored particles at the LHC have set a
limit of $\mgl \gsim 2 \tev$ (if not for very small $\mgl - \mneu1$),
placing a lower bound of $|M_1| \gsim 350 \gev$. Consequently, the
MSSM scenarios that can describe the soft-lepton excesses do not
follow the GUT-based mass pattern. 

In this paper we propose a scenario that can describe well the
soft-lepton excesses, follows the GUT-based mass patterns and is in
agreement with the searches for gluinos, the \sihi scenario:
the NMSSM with a singlino
dominated LSP, with $200 \gev \approx |\mu| < M_1 \sim M_2/2 \sim M_3/6$, 
and $\mgl \gsim 2 \tev$. This parameter space corresponds to
higgsino-like $\neu{2,3}$ and $\cha1$. The LHC searches are
interpreted as searches for
$pp \to \neu{2,3} \cha1 \to \neu1 Z^{*} \, \neu1 W^{\pm *}$ in the
NMSSM, leading to the desired 2~or 3~soft-lepton plus \ETmiss\ final
states.

We scanned over the relevant NMSSM parameter space, employing the GUT 
relation $M_1 \sim M_2/2 \sim M_3/6$, with $|\mu| \le M_1$ and
$M_1 \ge 350 \gev$. We have taken into
account all relevant experimental constraints: the DM relic density,
the DD cross section limits (the indirect detection limits do not play
a role), flavor observables, the LHC BSM
Higgs-boson searches, and the LHC Higgs-boson rate measurements
(where we identify the lightest $\cp$-even Higgs boson with the Higgs
discovered at the LHC). In particular, we apply 
the limits from the 2~and 3~soft-lepton searches (making
use of the NMSSM reinterpretation given in \citere{Agin:2024yfs}).
We demonstrated that the \sihi scenario can yield a perfect description
of the observed soft-lepton excesses, while being in agreement with
all existing constraint and fulfilling the GUT relations for~$M_{1,2,3}$. 

The preferred mass ranges are $\mneu2 \gsim 180 \gev$,
$\De m_{21} := \mneu2 - \mneu1 \approx 12-14 \gev$,
$\mneu3 - \mneu2 \approx 3-13 \gev$,
$\mcha1 - \mneu2 \approx 0-5 \gev$.
In the lower mass range the golden channel has a production cross
section in the order of one pb. 
The \sihi scenario furthermore predicts a second lightest $\cp$-even
Higgs boson in the range of $125 \lsim \mHz \lsim 225 \gev$ with a
singlet component larger than 0.9, which may yield interesting
prospects for the HL-LHC.
The \sihi scenario can yield the full DM content of the universe. The
future DD searches 
show interesting prospects for the next experimental runs, in particular for the spin independent
searches. However, for all types of DD searches (spin independent, spin dependent with protons or neutrons) we
find parameter points that are below the respective $\nu$-fog limits and may thus escape the searches 
(using the established experimental techniques).

We have also studied whether it would be possible to probe these scenarios via
the characterization of the $h_1$, i.e.\ the lightest $\cp$-even Higgs boson of the NMSSM, 
which is interpreted as the Higgs-boson discovered at the LHC.
A prime target for this approach is the decay of the $h_1 \to \ga\ga$.
We have found that, while non-negligible deviations from the SM are possible for the effective
coupling to two photons, there are regions of the preferred parameter space that
features experimentally indistinguishable deviation with respect to the SM. In other words,
while the observation of a small deviation in $h_1 \to \ga\ga$ can very well be explained
in our scenario, it will not be possible to use these measurements to completely test
it. On the other hand, the search for the other Higgs states of the NMSSM appear as an interesting
target. We leave such an analysis for a future study. 

\medskip
For the first time, CMS and ATLAS observe consistent excesses in the
search for SUSY particles. 
These excesses are observed in the processes
$pp \to \neu2\cha1 \to \neu1 Z^*\;\neu1 W^*$: 2~soft leptons plus
$\ETmiss$, and  3~soft leptons plus $\ETmiss$. 
While each experiment and search
individually is not significant by itself, the occurrence of excesses in
multiple search channels, observed by both CMS and ATLAS, gives rise to
the hope that indeed for the first time BSM physics has been observed.
The LHC Run~3 should give conclusive results, either refuting the
observed excesses, or discovering SUSY.

%%%%%%%%%%%%%%%%%%%%%%%%%%%%%%%%%%%%%%%%%%%%%%%%%%%%%%%%%%%%%%%%%%%%%%%%%%
%\clearpage

\subsection*{Acknowledgments}

We thank
M.~Martinez
for helpful discussions.
I.S.\ acknowledges support from~SERB-MATRICS, ANRF, India, under grant no. MTR/2023/000715.
The work of S.H.\ has received financial support from the
grant PID2019-110058GB-C21 funded by
MCIN/AEI/10.13039/501100011033 and by ``ERDF A way of making Europe", 
and in part by the grant IFT Centro de Excelencia Severo Ochoa
CEX2020-001007-S funded by MCIN/AEI/10.13039/501100011033. 
S.H.\ also acknowledges support from Grant PID2022-142545NB-C21 funded by
MCIN/AEI/10.13039/501100011033/ FEDER, UE. E.B.\ would like to thank the LNF INFN Data Cloud Team
and the CERN-IT department for the use of their computational resources.

%\clearpage
%\pagebreak

%%%%%%%%%%%%%%%%%%%%%%%%%%%%%%%%%%%%%%%%%%%%%%%%%%%%%%%%%%%%%%%%%%%%
%%%%%%%%%%%%%%%%%%%%%%%%%%%%%%%%%%%%%%%%%%%%%%%%%%%%%%%%%%%%%%%%%%%%

\newcommand\jnl[1]{\textit{\frenchspacing #1}}
\newcommand\vol[1]{\textbf{#1}}

\end{document}